\documentclass[a4paper]{aa}
\usepackage{lscape}

\usepackage{graphicx} 
\usepackage{latexsym,amsmath,amssymb}
\usepackage{natbib}
\bibpunct{(}{)}{;}{a}{}{,}

\def \suzaku {\it Suzaku}
\def \swift {\it Swift}
\def \nustar {\it NuSTAR}

\def \hcm {\hbox {\ifmmode $ atom cm$^{-2}\else atom cm$^{-2}$\fi}}
\def \arcmin {\hbox{$^\prime$}}
\def \arcsec {\hbox{$^{\prime\prime}$}}
\def \deg {$^{\circ}$}

\def \rchisq {$\chi_{\nu} ^{2}$}
\def \approxgt{\mathrel{\hbox{\rlap{\lower.55ex \hbox {$\sim$}}
        \kern-.3em \raise.4ex \hbox{$>$}}}}
\def \approxlt{\mathrel{\hbox{\rlap{\lower.55ex \hbox {$\sim$}}
        \kern-.3em \raise.4ex \hbox{$<$}}}}

\def \srcone {Aql\,X--1}

\def \maxi {MAXI\,J1836$-$194}

\newcommand {\ktdbb} {$kT_{\rm dbb}$}
\newcommand {\ktbb} {$kT_{\rm bb}$}

\def \nhabs {$N{\rm _H^{abs}}$}

\def \kbb {$k_{\rm bb}$}
\def \kdbb {$k_{\rm dbb}$}

\begin{document}

\title{The evolving jet spectrum of the neutron star X-ray binary Aql~X-1 in transitional states during its 2016 outburst}

\author{M. D{\'i}az Trigo\inst{1} \and D. Altamirano\inst{2} \and T. Din\c{c}er\inst{3} \and J.~C.~A. Miller-Jones\inst{4} \and D.~M. Russell\inst{5} \and A. Sanna\inst{6} \and C. Bailyn\inst{3} \and F. Lewis\inst{7,8} \and S. Migliari\inst{9,10} \and F. Rahoui\inst{11}
}
\institute{
ESO, Karl-Schwarzschild-Strasse 2, D-85748 Garching bei M\"unchen, Germany
\and
Department of Physics \& Astronomy, University of Southampton, Southampton, Hampshire SO17 1BJ, UK
\and
Department of Astronomy, Yale University, P.O. Box 208101, New Haven, CT 06520-8101, USA
\and
International Centre for Radio Astronomy Research, Curtin University, GPO Box U1987, Perth,
Western Australia 6845, Australia
\and
New York University Abu Dhabi, PO Box 129188, Abu Dhabi, UAE              
\and
Dipartimento di Fisica, Universit\`a degli Studi di Cagliari, SP Monserrato-Sestu km 0.7, I-09042 Monserrato, Italy
\and
Faulkes Telescope Project, School of Physics and Astronomy, Cardiff University, 5 The Parade, Cardiff, CF24 3AA, Wales, UK
\and
Astrophysics Research Institute, Liverpool John Moores University, 146 Brownlow Hill, Liverpool L3 5RF, UK
\and
XMM-Newton Science Operations Centre, ESAC/ESA, Camino Bajo del Castillo s/n, Urb. Villafranca del Castillo, 28691 Villanueva de la Ca\~nada, Madrid, Spain
\and
Institute of Cosmos Sciences, University of Barcelona, Mart\'i i Franqu\`es 1, 08028 Barcelona, Spain
\and
Department of Astronomy, Harvard University, 60 Garden street, Cambridge, MA 02138, USA
}

\date{Received ; Accepted:}

\authorrunning{D{\'i}az Trigo et al.}

\titlerunning{ }

\abstract{We report on quasi-simultaneous observations from radio to X-ray frequencies of the neutron star X-ray binary \srcone\ over accretion state transitions during its 2016 outburst. All the observations show radio to millimetre spectra consistent with emission from a jet, with a spectral break from optically thick to optically thin synchrotron
emission that decreases from $\sim$\,100~GHz to $<$\,5.5~GHz during the transition from a hard to a soft accretion state. The 5.5~GHz radio flux density as the source reaches the soft state, 0.82\,$\pm$\,0.03~mJy, is the highest recorded to date for this source. During the decay of the outburst, the jet spectral break is detected again at a frequency of $\sim$~30--100~GHz. The flux density is 0.75\,$\pm$\,0.03~mJy at 97.5~GHz at this stage. This is the first time that a change in the frequency of the jet break of a neutron star X-ray binary has been measured, indicating that the processes at play in black holes are also present in neutron stars, thus supporting the idea that the internal properties of the jet rely most critically on the conditions of the accretion disc and corona around the compact object, rather than the black hole mass or spin or the neutron star surface or magnetic field.  
  
} \maketitle

\section{Introduction}
\label{sec:intro}

Relativistic jets from supermassive black holes (BHs) at the centres of galaxies and stellar-mass BHs in X-ray binaries (XRBs) have been extensively studied for decades. However, key questions including 
where and why they are formed, how they are collimated and whether the energy comes from the accretion disc or the spin of the BH remain unanswered. Observationally, it has been established that the presence/absence of a compact or transient jet in BH XRBs is related to the  accretion flow geometry during a transient outburst \citep{fender04mnras}. A steady jet is present from the beginning of the outburst, when the source is in the so-called ``hard" state. The jet persists while the luminosity rises and until the X-ray spectrum starts to soften. During the transition to the ``soft" state the steady jet emission is quenched \citep[see e.g.][]{russell11apj, 1743:coriat11mnras} and in some sources discrete ejecta are launched and their radio emission can be spatially resolved \citep{1915:mirabel94nature,1655:tingay95nature}, although these ejections are not found in every transition \citep[][]{1659:paragi13mnras}. When the BH transitions back to the hard state at a lower luminosity, the compact jet is reinstated \citep{kalemci05apj,kalemci13apj}.

In the last years, major efforts have been devoted to understanding what determines the frequency of the jet spectral break from optically thick to optically thin emission \citep{gx339:gandhi11apjl,1659:horst13mnras,gx339:corbel13bmnras,1836:russell13mnras,1836:russell14mnras,koljonen15apj}. This break together with the cooling break expected at higher energies \citep{sari98apj} are proxies for the total radiative power of the jet and may provide important information for understanding jet launching and quenching. \citet{1836:russell13mnras,1836:russell14mnras} performed multi-wavelength observations of the BH XRB \maxi\ during its 2011 outburst. They found that the spectral break from optically thick to optically thin emission moved from infrared (IR) to radio frequencies as the spectrum got softer at a fairly constant luminosity during a spectral state transition and then moved back up again as the spectrum got harder at the end of the outburst. This indicates that the main player in determining the break frequency is the changing structure of the accretion flow, rather than the mass and spin of the BH or its luminosity. 

The temporal evolution of the optically thick-to-thin jet break has never been studied over the course of a neutron star (NS) XRB outburst. In fact, studies of NS jets have so far been relatively sparse, due to their faintness in the radio band \citep[see][and references therein]{migliari06mnras}. However, such studies are of fundamental importance since they offer the possibility to isolate the role played by the BH event horizon or the NS surface or magnetic field in powering the jets.

\srcone\ is a NS \citep{aql:koyama81apj} low-mass X-ray binary (LMXB) undergoing accretion outbursts every 200--300 days. Its distance was determined from type I X-ray bursts to be 4.5--6 kpc \citep[][]{aql:jonker04mnras}\footnote[1]{Hereafter, we adopt a distance of 5.2~kpc when calculating luminosities.}. Based on its spectral and timing properties, it was classified as an atoll-type X-ray binary \citep{aql:reig04apj}. \citet{aqlx1:casella08apj} observed millisecond X-ray pulsations with a frequency of 550.27~Hz that were identified with the NS spin frequency. The pulsations only lasted for 150~s and were never detected again, highlighting the possibility of transient pulsations in any of the ``non--pulsating" XRBs. The optical counterpart of \srcone\ was identified as a K7V star by \citet{aql:chevalier99aa}. Recently, \citet{aql:matasanchez17mnras} used the Very Large Telescope (VLT)/SINFONI to obtain time resolved spectra for both \srcone\ and the interloper star located less than 0.5\arcsec\ from \srcone, and classified the optical counterpart to \srcone\ as a K4\,$\pm$\,2 star moving at a projected velocity of 136\,$\pm$\,4~km s$^{-1}$. They determined its orbital inclination to be 36--47\deg, lower than that of 72--79\deg\ determined by  \citet{aql:galloway16mnras} based on the detection of two dipping episodes in the {\it RXTE} X-ray light curves. The orbital period of the binary was determined to be 19 hours \citep{aql:chevalier91aa, aql:welsh00aj} from the light curve modulations in quiescence.  

Radio emission from \srcone\ was first detected with the Very Large Array (VLA) during its 1990 outburst with flux densities of 0.4 and 0.13~mJy \citep{aqlx1:hjellming90iauc}. \citet{aqlx1:tudose09mnras} analysed the March 2002 outburst and did not find any significant correlation between the radio and optical (R--band) emission, consistent with \citet{aqlx1:maitra08apj} who attributed the optical-infrared (OIR) emission to thermal heating of the irradiated outer accretion disc. \citet{aql:miller-jones10apj} investigated the radio--X-ray coupling during the entire 2009 outburst and found that in the hard state it is qualitatively similar to that found for BHs, in that the radio and X-ray fluxes positively correlate. Moreover, they found that during the soft state the radio emission was quenched to $\approxlt$\,0.2~mJy for luminosities $\approxgt$\,10\% Eddington.

\citet{aqlx1:maitra08apj} first classified the outbursts of \srcone\ in two types depending on the OIR light curve 
morphology: the fast rise and exponential decay (FRED) type outburst seen in many soft X-ray transients, and the low-intensity 
state (LIS), where the optical-to-soft X-ray flux ratio is much higher than during a FRED. They did not find a single correlation
 between the optical (R--band) and soft X-ray fluxes suggesting that those two types of outbursts have fundamentally different 
 accretion flow properties. The OIR light curves did not show any signature of compact jets or accretion state changes. According to 
this classification, between June 1998 and November 2007 only three outbursts were included in the FRED-only category, namely 
those that occurred on October 2000, March 2002 and March 2003. 

\citet{aqlx1:asai12pasj} also classified the \srcone\ outbursts but based on the relative intensity evolutions below/above 15~keV. They identified slow (S-type) and fast (F-type) outbursts, the former having a longer initial hard state ($\approxgt$~9~d) than the latter ($\approxlt$~5~d). Importantly, for the S-type, the 15--50~keV intensity reaches the maximum before the hard-to-soft transition, while the opposite is true for the F-type (see their Fig.~8). They also found that the S-type outbursts had higher luminosity both before the start of the outburst and at the hard-to-soft state transition and concluded that the X-ray irradiation during the pre-outburst period would determine whether the outburst becomes S-type or F-type. In the S-type, the hard-to-soft state transition occurred at $\sim$5\% of the Eddington luminosity. Examples of S-type outbursts in this classification occurred on November 2009, September 2010 or October 2011. 

Finally, \citet{aqlx1:gungor14mnras} made a slightly different classification based on the duration and maximum flux of the X-ray emission and identified three types of outbursts: long-high (including the October 2000 FRED outburst and the 2011 S-type outburst), medium-low (February 2002 and September 2010) and short-low (November 2009). They also concluded that irradiation is the dominant physical process that 
leads to the different classes of outbursts. 

Following the exploratory Atacama Large Millimeter/submillimeter Array (ALMA) study of NSs by \citet{1820:diaz17aa}, we successfully proposed for quasi-simultaneous observations of a NS transient at radio, millimetre (mm), IR and X-ray wavelengths. The goal was to study the behaviour of the jet break over the transitional states for the first time in a NS LMXB and to test the hypothesis that the jet break moves to lower frequencies as the source evolves towards softer states, as was observed for the BH LMXB \maxi\ \citep{1836:russell13mnras,1836:russell14mnras}. In NSs, jet spectral breaks lying in the IR band have been reported in three sources \citep{0614:migliari10apj,baglio16aa,1820:diaz17aa} but we expect large changes in the frequency of the jet break during state transitions, based on the behaviour of \maxi.
We triggered our multi-wavelength campaign in August and September 2016, during a recent outburst of \srcone. In this paper, we report on the results of this campaign. 

\section{Observations and data analysis}
\label{sec:observations}

We observed \srcone\ during August and September 2016 at radio, mm, IR, optical and X-ray wavelengths with the Australia Telescope Compact Array (ATCA), ALMA, the VLT, the SMARTS 1.3~m telescope, the Faulkes telescopes, Las Cumbres Observatory (LCO) telescopes and $\swift$ \citep{swift:gehrels04apj}, respectively. We also used one contemporaneous $\nustar$ observation for analysis. Table~\ref{tab:obslog} shows a log of the radio, mm, mid-IR and $\swift$ X-ray pointed observations. 

\begin{figure*}[ht]
\hspace{-1cm}
\includegraphics[angle=180.0,width=0.85\textheight]{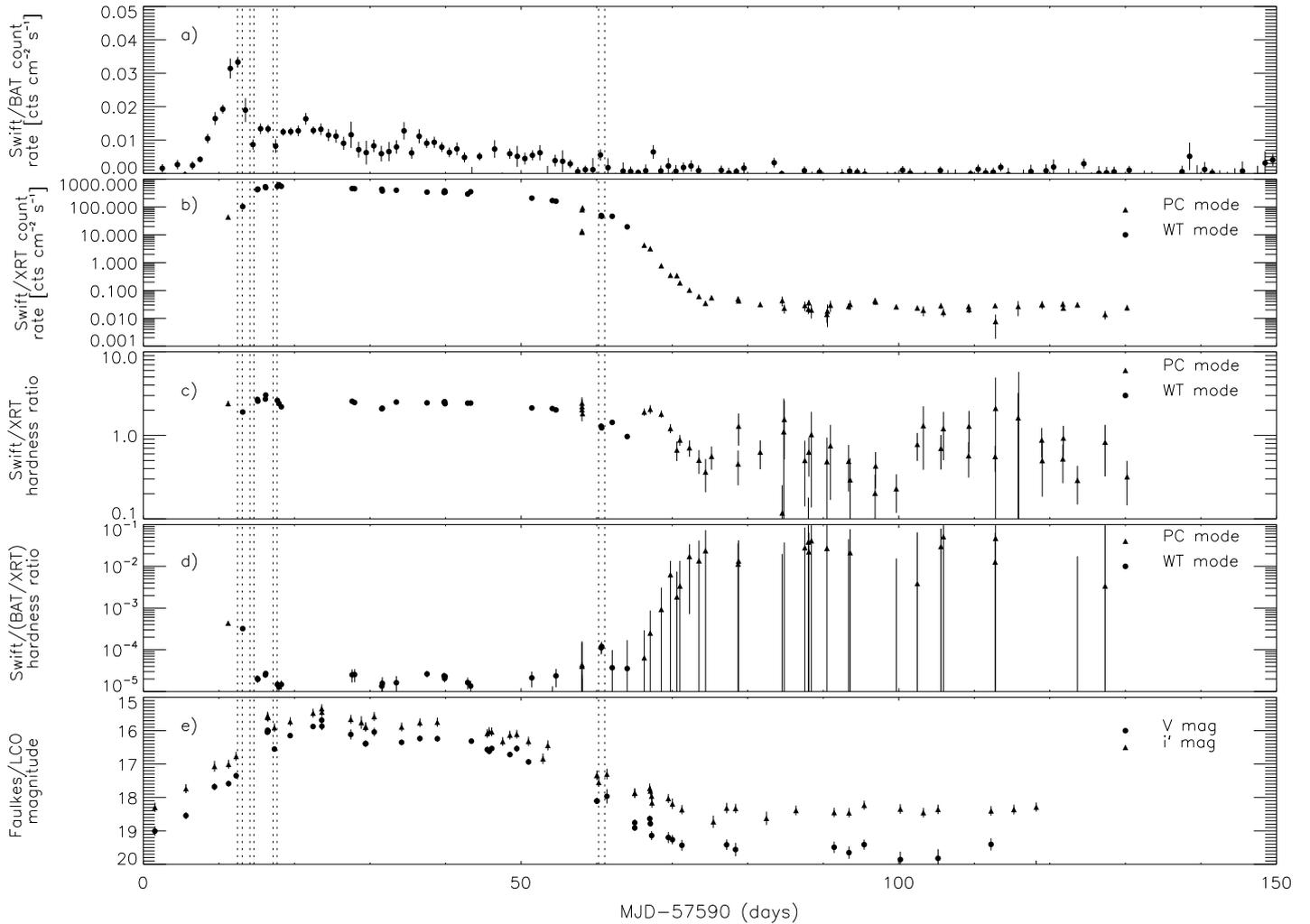}
\caption{X-ray and optical light curves of \srcone. The start and end times of the four multi-wavelength observations listed in Table~\ref{tab:obslog} are indicated with dotted lines. The start of each multi-wavelength observation is defined as the start time of the first observation within a set, e.g. the start of the ATCA/VLT observations mark the start of obs~1/2a, respectively. Similarly, the end of each multi-wavelength observation is defined as the end time of the last observation within a set, e.g. the end of the $\swift$/ALMA observations mark the end of obs~1/2a, respectively.
a) $\swift$/BAT 15--50 keV count rate. b) $\swift$/XRT 0.3--10 keV count rate (count rates in PC mode could be affected by pile-up at count rates above $\sim$\,1~cts~s$^{-1}$). c) $\swift$/XRT hardness ratio (1.5--10 keV/0.3--1.5 keV count rate). d) $\swift$/(BAT/XRT) hardness ratio (15--50 keV/0.3--10 keV count rate). The hardness ratio is calculated for days on which there is a $\swift$/XRT point and a $\swift$/BAT point within one day. e) Faulkes/LCO V and $i\arcmin$ magnitudes. 
}
\label{fig:lc}
\end{figure*}

Fig.~\ref{fig:lc} shows the light curve of the 2016 outburst as covered by $\swift$ and the Faulkes/LCO optical telescopes including the times of the radio, mm and mid-IR pointed observations presented here. To generate the $\swift/XRT$ light curves we estimated the average-per-observation intensity (count rate in the  0.3--10 keV band) and hardness ratio (HR, ratio of the count rates in the 1.5--10 keV and 0.3--1.5 keV energy bands)  using the online light curve generator provided by the UK Swift Science Data Centre \citep[UKSSDC;][]{evans07aa}. The $\swift/BAT$ data are provided by the 
$\swift$ Team as daily averages in the energy range 15--50 keV \citep{krimm13apjs}.

\begin{table}
\begin{center}
\caption[]{Observation log. The single observations within each subset are ordered by increasing frequency.}
\begin{tabular}{l@{\extracolsep{1mm}}l@{\extracolsep{1mm}}l@{\extracolsep{1mm}}l@{\extracolsep{1mm}}}
\hline \noalign {\smallskip}
Observatory & Start time (UT) & End time (UT) &  \\
\hline \noalign {\smallskip}
Obs~1 \\
\noalign {\smallskip}
ATCA 5.5 \& 9 GHz & 02 Aug 2016 10:34 & 02 Aug 2016 15:30 & \\
ALMA 97.49 GHz & 03 Aug 2016 01:39 & 03 Aug 2016 02:06 & \\
ALMA 302.99 GHz & 03 Aug 2016 00:35 & 03 Aug 2016 01:38 & \\
VLT/VISIR & 03 Aug 2016 00:27 & 03 Aug 2016 01:26 \\
$\swift$/XRT & 03 Aug 2016 03:03 & 03 Aug 2016 03:30 \\
\smallskip\\
Obs~2a \\
\noalign {\smallskip}
ATCA 5.5 \& 9 GHz & 04 Aug 2016 09:27 & 04 Aug 2016 14:59 \\
ALMA 302.99 GHz & 05 Aug 2016 05:47 & 05 Aug 2016 06:51 & \\
VLT/VISIR & 04 Aug 2016 02:20 & 04 Aug 2016 04:15 \\
$\swift$/XRT & 05 Aug 2016 01:53 & 05 Aug 2016 03:27 \\
\smallskip\\
Obs~2b \\
\noalign {\smallskip}
ATCA 5.5 \& 9 GHz & 07 Aug 2016 12:16 & 07 Aug 2016 16:30 \\
ALMA 97.49 GHz & 07 Aug 2016 03:49 & 07 Aug 2016 04:19 & \\
$\swift$/XRT & 07 Aug 2016 17:18 & 07 Aug 2016 17:34 \\
$\nustar$ & 07 Aug 2016 11:04 & 08 Aug 2016 05:18 \\ 
\smallskip\\
Obs~3 \\
\noalign {\smallskip}
ATCA 5.5 \& 9 GHz & 19 Sep 2016 06:00 & 19 Sep 2016 11:00 & \\
ALMA 97.49 GHz & 19 Sep 2016 22:10 & 19 Sep 2016 22:36 & \\
ALMA 302.99 GHz & 20 Sep 2016 00:32 & 20 Sep 2016 01:35 & \\
$\swift$/XRT & 19 Sep 2016 14:10 & 19 Sep 2016 14:29 \\
\noalign {\smallskip} \hline 
\label{tab:obslog}
\end{tabular}
\end{center} 
\end{table}

\begin{table*}
\begin{center}
\caption[]{Fluxes, flux densities and/or count rates of the observations reported. For $\swift$/BAT we list either the orbital count rate closest in time to the $\swift$/XRT observation within a given day, or the daily count rate ({\it d}) if no $\swift$/XRT observations were available on the same day. For SMARTS/LCO the errors are typically smaller than 0.1 magnitudes, making the flux errors typically smaller than 10$\%$ and than the symbols in Fig. \ref{fig:sed}. Upper limits are given at a 3$\sigma$ significance level.}
\begin{scriptsize}
\begin{tabular}{l@{\extracolsep{1mm}}cl@{\extracolsep{1mm}}l@{\extracolsep{1mm}}l@{\extracolsep{1mm}}l@{\extracolsep{1mm}}l@{\extracolsep{1mm}}l@{\extracolsep{1mm}}l@{\extracolsep{1mm}}l@{\extracolsep{1mm}}l@{\extracolsep{1mm}}}
\hline \noalign {\smallskip}
Date & MJD-- & \multicolumn{2}{l}{ATCA} & \multicolumn{2}{l}{ALMA} &   \multicolumn{2}{l}{VLT/VISIR} & $\swift$/XRT & $\swift$/BAT \\
& 57590 & 5.5 GHz & 9 GHz & 97.49 GHz & 302.99 GHz & \multicolumn{2}{l}{8.7--11.5 $\mu$m} & 0.4--9~keV & 15--50~keV \\
& & [$\mu$Jy] & [$\mu$Jy] & [$\mu$Jy] & [$\mu$Jy]  & [mJy]  & [mJy] & [10$^{-9}$ erg cm$^{-2}$ s$^{-1}$] & [cts cm$^{-2}$ s$^{-1}$] \\
\hline \noalign {\smallskip}
02 Aug 2016 & 12 & 515\,$\pm$\,26 & 428\,$\pm$\,28 & -- & -- & -- & -- & -- & 0.033\,$\pm$\,0.002 ({\it d}) \\
03 Aug 2016 & 13 &  --  &  --  & 636\,$\pm$\,24 & 532\,$\pm$\,42 & $<$\,1.603 (B10.7) & -- & 5.60\,$\pm$\,0.04 & 0.025\,$\pm$\,0.004 \\
\noalign {\smallskip}
04 Aug 2016 & 14 & 494\,$\pm$\,19 & 208\,$\pm$\,18 & -- & -- & $<$\,1.931 (B11.7) & $<$\,1.820 (J8.9) & -- & 0.009\,$\pm$\,0.002 ({\it d}) \\
05 Aug 2016 & 15 & -- &  -- & -- & $<$189 & -- & -- &  23.1\,$\pm$\,0.1 & 0.014\,$\pm$\,0.004 \\
\noalign {\smallskip}
07 Aug 2016 & 17 & 810\,$\pm$\,19 & 528\,$\pm$\,19 & 406\,$\pm$\,26 & -- & -- &  -- & 28.6\,$\pm$\,0.3 & 0.011\,$\pm$\,0.004 \\
\noalign {\smallskip}
19 Sep 2016 & 60 & 259\,$\pm$\,13 & 366\,$\pm$\,12 & 753\,$\pm$\,29 & -- & -- & -- & 1.94\,$\pm$\,0.02 & 0.009\,$\pm$\,0.003\\
20 Sep 2016 & 61 & -- & -- & -- & 598\,$\pm$\,63 & -- & -- & -- & 0.002\,$\pm$\,0.003 ({\it d})\\
\noalign {\smallskip} 
\noalign {\smallskip} \hline
\noalign {\smallskip} 
 & & SMARTS H & SMARTS J & LCO Y& SMARTS I & LCO i$\arcmin$ & SMARTS/LCO R & SMARTS/LCO V & LCO B  \\
&  & $1.650~\mu$m & $1.250~\mu$m & $1.004~\mu$m & $0.798~\mu$m & $0.755~\mu$m & $0.641~\mu$m & $0.545~\mu$m & $0.436~\mu$m \\
& & \multicolumn{7}{c}{[mJy]} & \\
\hline \noalign {\smallskip}
02 Aug 2016 & 12 &  -- &  -- &  -- &  -- & 1.82 & 1.79 & 1.71 & 1.63 \\
03 Aug 2016 & 13 &  -- & 1.69 & --  &  -- &   -- & 1.82 & 1.86 &   --\\
\noalign {\smallskip}
04 Aug 2016 & 14 &  -- & 2.40 &  -- & 3.41 &  -- & 4.02 & 4.66 & -- \\
05 Aug 2016 & 15 &  -- &  -- &  -- &  -- &  -- &  -- & 4.66 & -- \\
\noalign {\smallskip}
07 Aug 2016 & 17 & 2.07 & 3.08 & 3.86 & 3.47 & 4.01 & 3.53 & 3.97 & 3.02\\ 
\noalign {\smallskip}
19 Sep 2016 & 60 & 0.41 & 0.68 & 0.65 & 0.82 & 0.89 & 0.78 & 0.86 & 0.61\\
\noalign {\smallskip} \hline 
\label{tab:fluxlog}
\end{tabular}
\end{scriptsize}
\end{center} 
\end{table*}

\subsection{ATCA}
\label{sec:ATCA}

We made four ATCA observations of Aql X-1, on the 2nd, 4th and 7th of August and the 19th of September, under program code C3010 (see Table~\ref{tab:obslog}). All observations were taken simultaneously in two 2-GHz frequency bands, centred at 5.5 and 9.0 GHz.   
The three observations in August were taken in the very compact H75 array configuration, and the September observation was made in the slightly more extended H168 configuration. These configurations are both hybrid arrays, with maximum baselines for the inner five antennas of 89 and 192\,m, respectively, and the sixth antenna located 4.4~km away and of little use for direct imaging. Thus, these compact configurations provide relatively poor angular resolution, but have the advantage of some north-south resolution due to the location of two antennas on the north spur.  This makes imaging an equatorial source such as Aql X-1 much more reliable than would be the case for an east-west interferometer such as the ATCA in a more extended configuration. However, one of the inner antennas was missing in each of the four observations, 
reducing the quality and sensitivity of our imaging.
We used 1934-638 as an amplitude and bandpass calibrator, and the nearby compact source 1849+005 to calibrate the antenna complex gains.  Data reduction was carried out using standard procedures within the Multichannel Image Reconstruction, Image Analysis and Display \citep[MIRIAD;][]{Sault1995} software package, and then imported into the Common Astronomy Software Application package \citep[CASA;][]{McMullin2007} for imaging.

The August observations had angular resolutions of $\sim 120\arcsec\times60\arcsec$ at 5.5\,GHz, and $\sim 75\arcsec\times40\arcsec$ at 9.0\,GHz, improving to $\sim 55\arcsec\times27\arcsec$ and $\sim 33\arcsec\times18\arcsec$, respectively, in September. Due to the poor angular resolution and the sparse $uv$-coverage we undertook a number of steps to extract the most accurate flux densities for \srcone. Firstly, we subtracted a confusing source 43.7\,\arcsec\ west of Aql X-1, at (J2000) 19:11:13.40 $+00$:34:47.7, with a flux density of 240\,$\mu$Jy at 5.25\,GHz and a spectral index of $\alpha=-1.08$, which was detected in higher-resolution archival VLA observations (A. Deller, private communication) but due to the compact array configuration was not resolved in our images, except for the September observation at 9\,GHz. For this observation we measured the flux density of the confusing source to be 134$\pm$15\,$\mu$Jy, fully consistent with the 134\,$\mu$Jy expected at 9\,GHz from the measured VLA spectrum. Thus, we confirm that the confusing source is not significantly variable and that the procedure used is adequate to fully remove it from our data. Secondly, we created an image of the field stacking all four epochs and subtracted all  sources out to a distance of 17\,\arcmin\ from the phase centre in the $uv$-plane after refining their positions using the NRAO VLA Sky Survey (NVSS) catalogue where available \citep{Condon98}.
Finally, we extracted flux densities using two methods: the conventional image-plane analysis, and directly fitting the target in the $uv$-plane using the {\sc uvmultifit} algorithm of \citet{marti-vidal14}, expected to be more reliable when the $uv$-coverage is poor. After checking that the values were broadly consistent for both methods, we adopted the values obtained with the second method (the $uv$-plane fitting) and list them in Table~\ref{tab:fluxlog}. 

Aql X-1 was detected at all four ATCA epochs, at both 5.5 and 9.0\,GHz.  The emission initially dropped between the 2nd and 4th of August, before peaking on the 7th of August at $810\pm19$\,$\mu$Jy at 5.5\,GHz and $528\pm19$\,$\mu$Jy at 9.0\,GHz.

\subsection{ALMA}

ALMA observed \srcone\ four times from August to September 2016 (see Table~\ref{tab:obslog}). All observations were set up to use four spectral windows, each with a bandwidth of 2 GHz. The spectral windows were centred at 90.49, 92.43, 102.49 and 104.49~GHz for the three observations in band 3 and at 295.99, 297.93, 307.99 and 309.99~GHz for the observations in band 7. The spectral resolution for each spectral window was 31.2 MHz.
The observations were performed using between 38 and 46 12-m antennas. The maximum baseline length was 1.4 and 3.1 km, for the August and September observations, respectively. The bandpass calibrator was J1751+0939 for all band 3 observations and the first band 7 observation and J1924-2914 for the second and third band 7 observations. The bandpass calibrator was also used as flux calibrator for all observations except the second band 3 observation, for which J1733-1304 was used. The phase calibrator was J1851+0035 for all observations except the last band 3 observation, for which J1907+0127 was used.  

We re-generated images from the ALMA calibrated products following standard procedures with CASA v4.7. For the band 7 observations, the images extracted with the initial calibrated products showed slightly extended emission. This could be due to loss of phase coherence since the weather conditions were marginal for observations at frequencies $\approxgt$\,300~GHz. Therefore, we performed a detailed check of the calibrated products and found significant phase variability within a calibrator scan for antennas further than 260, 400 and 400~m from the centre of the array for obs 1, 2a and 3, respectively. We re-calibrated the datasets for these observations flagging out the problematic antennas before re-generating the images. These images show point-like source emission, similarly to the band 3 observations. Notably, the measured flux densities increased by $\approxgt$50\% relative to the original pipeline-calibrated images, highlighting the improvement in the quality of the images. Consequently, we consider the images and flux densities obtained after re-calibration hereafter.

\begin{figure*}[ht]
\hspace{0cm}\includegraphics[angle=0.0,width=0.25\textheight]{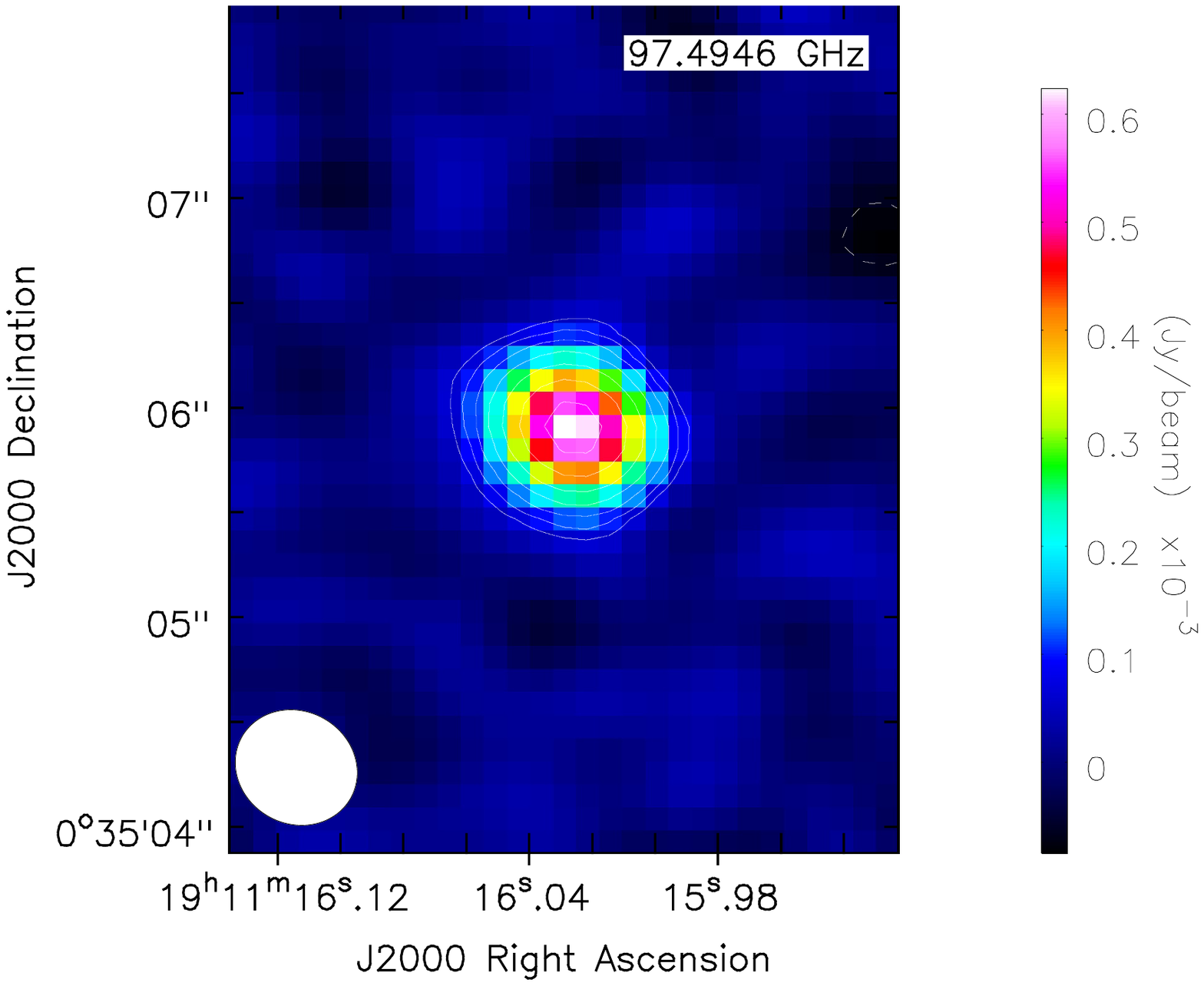}
\hspace{0cm}\includegraphics[angle=0.0,width=0.25\textheight]{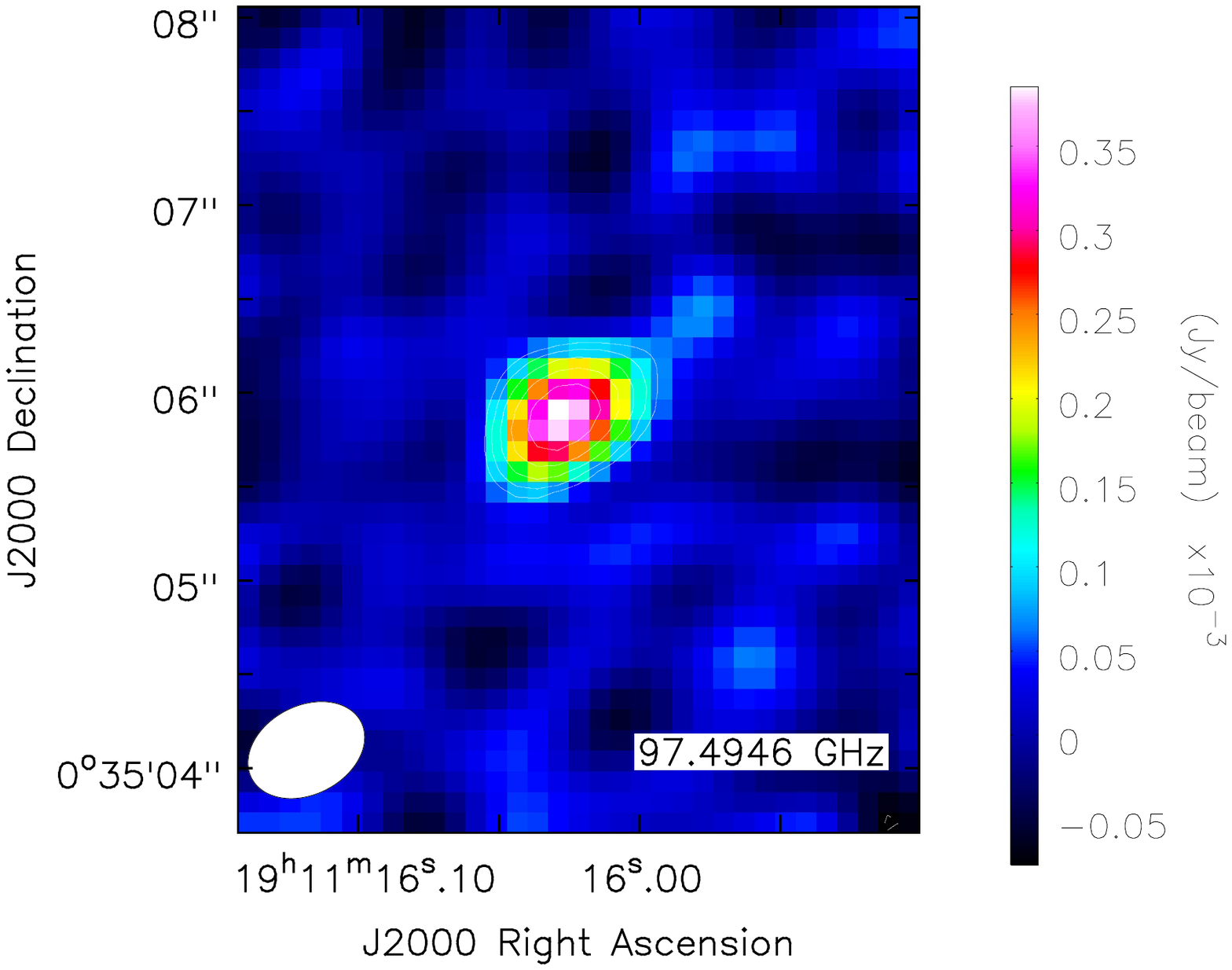}
\hspace{0cm}\includegraphics[angle=0.0,width=0.25\textheight]{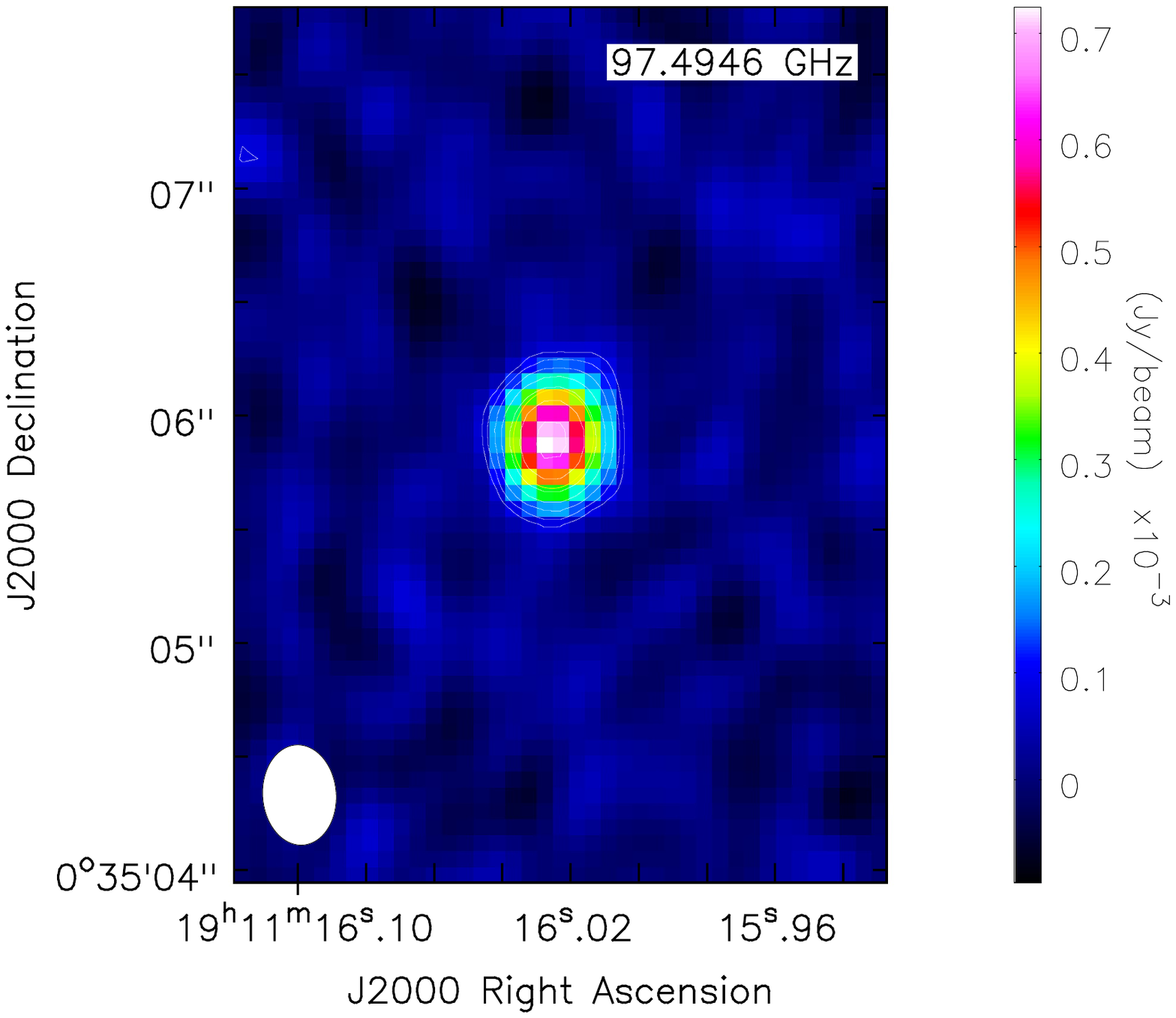}
\vspace{-4cm}
\hspace{0cm}\includegraphics[angle=0.0,width=0.265\textheight]{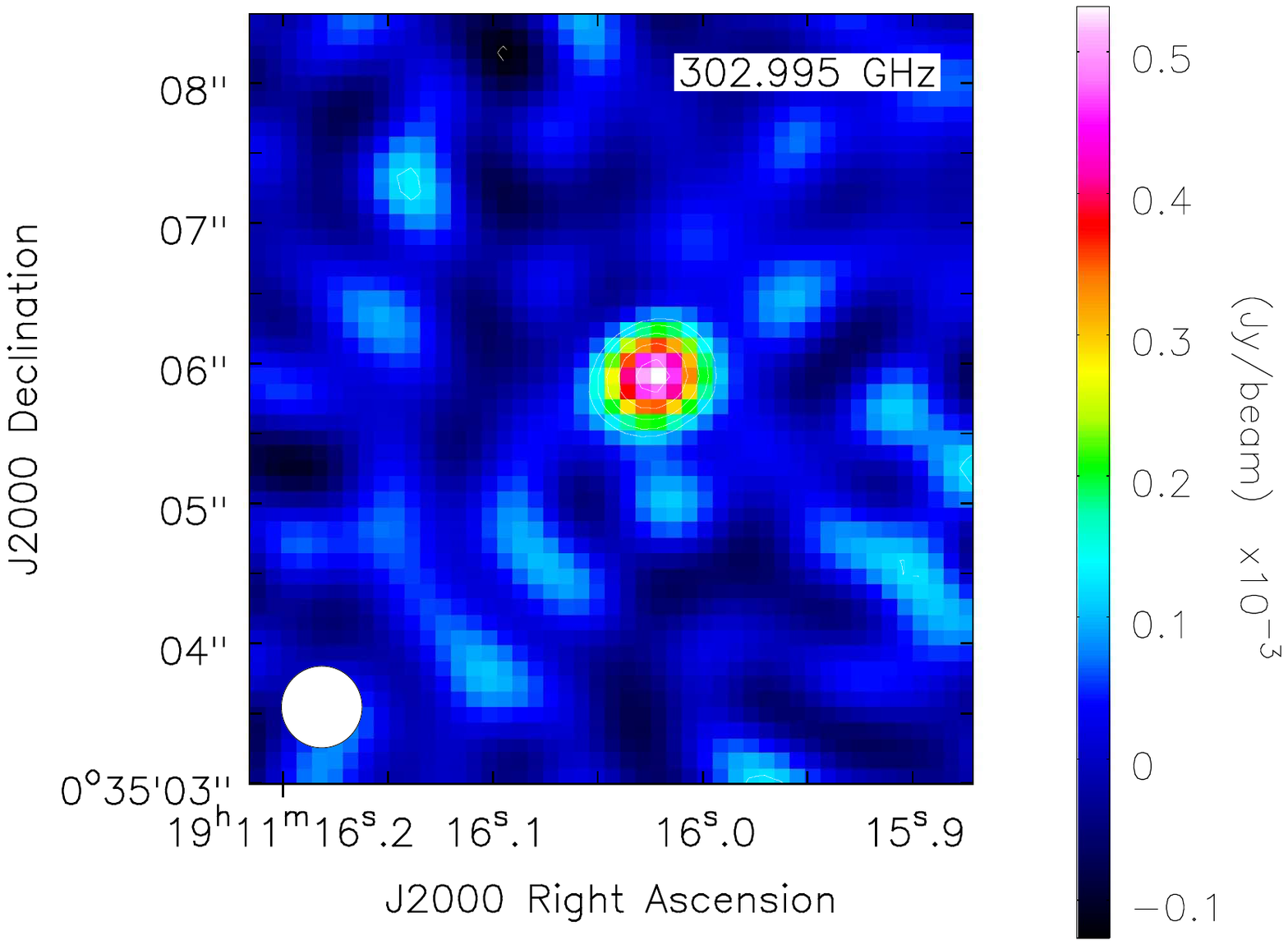}
\hspace{-0.4cm}
\hspace{0cm}\includegraphics[angle=0.0,width=0.265\textheight]{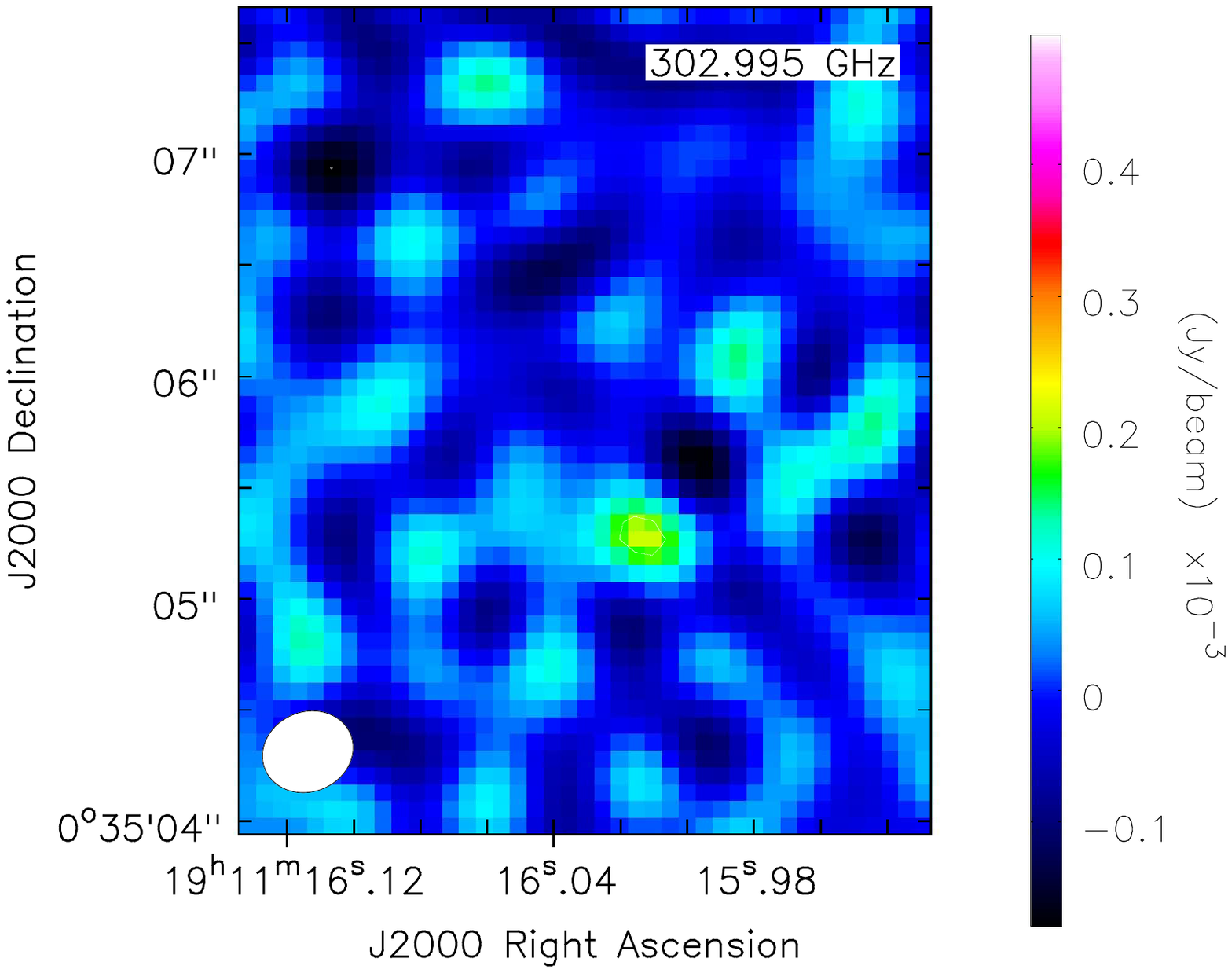}
\hspace{0cm}\includegraphics[angle=0.0,width=0.265\textheight]{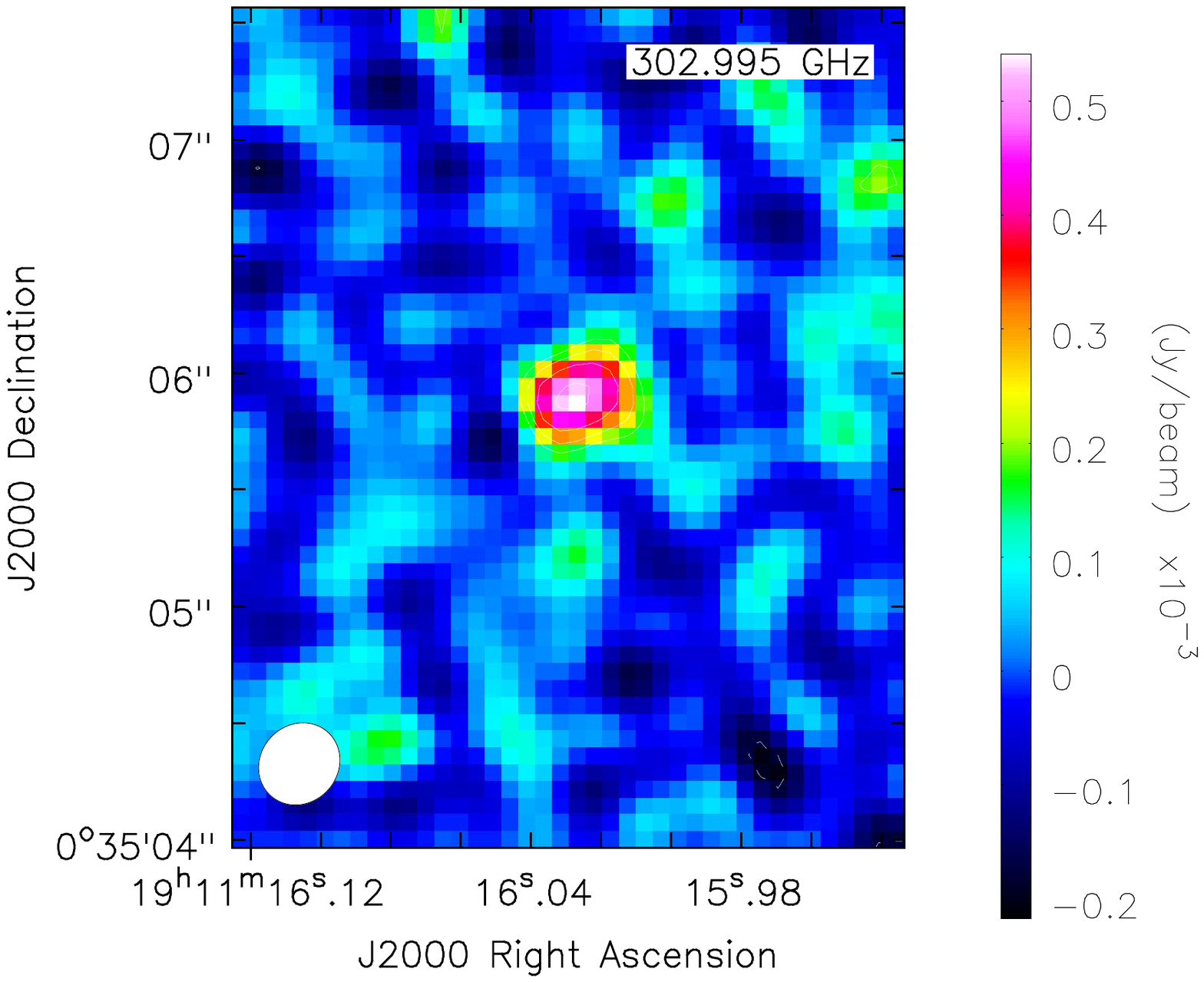}
\caption{ALMA images of the averaged data of \srcone\ at 97.49~GHz (upper panels) and 302.99~GHz (lower panels) during epochs 1--3 (from left to right). Contours are overlaid in the zoomed image at levels of $\pm (\sqrt 2)^n$ $\sigma$ significance with $n$ = 3, 4, 5...15 with respect to the rms of each observation (see Table~\ref{tab:fluxlog}). The synthesized beam is shown at the lower-left corner of each image. 
}
\label{fig:alma_images}
\end{figure*}

Fig.~\ref{fig:alma_images} shows the final reconstructed images at the position of \srcone, extracted with Briggs weighting and robust=0.5 for all the observations and Table~\ref{tab:fluxlog} lists the flux densities for all observations. 

At 97.49~GHz (band 3), the synthesised beam size was between 0.46\arcsec$\times$0.35\arcsec and 0.71\arcsec$\times$0.50\arcsec, depending on the epoch. The peak of radio emission is centred on average at (J2000) 19:11:16.024\,$\pm$\,0.005 +00:35:05.900\,$\pm$\,0.007 and the peak flux densities are 636\,$\pm$\,24, 406\,$\pm$\,26 and 753\,$\pm$\,29~$\mu$Jy for epochs 1, 2b and 3, respectively. The reported flux density for all sources in this section corresponds to the peak flux from the elliptical Gaussian fitted to the source emission in the image plane and the error corresponds to the rms of the image measured far away from the source position. 

At 302.99~GHz (band 7), the synthesised beam size was between 0.37\arcsec$\times$0.33\arcsec\ and 0.58\arcsec$\times$0.58\arcsec, depending on the epoch. The peak of radio emission is centred on average at (J2000) 19:11:16.02\,$\pm$\,0.01 +00:35:05.89\,$\pm$\,0.01 and the peak flux density is 532\,$\pm$\,42 and 598\,$\pm$\,63~$\mu$Jy for obs~1 and 3, respectively. During obs~2a, we do not detect the source and the 3$\sigma$ upper limit is $<$~189\,$\mu$Jy.

\subsection{VLT/SMARTS/LCO}
\label{sec:OIR}

Mid-IR observations of the field of \srcone\ were made with the VLT on the 3rd and 4th of August 2016 under program 097.D-0803 (PI: Russell). The VLT Imager and Spectrometer for the mid-IR \citep[VISIR;][]{LagageVISIR} instrument on UT3 (Melipal) was used in small-field imaging mode, with a pixel scale of 45 mas pixel$^{-1}$. These and more observations near the end of September will be fully presented in Russell et al. (in preparation), which focusses on the evolution of the mid-IR to UV spectral energy distributions (SEDs) in detail. The August observations were performed in the B10.7 (centred at $10.65 \mu$m) filter on the 3rd of August and the J8.9 ($8.72 \mu$m) and B11.7 ($11.52 \mu$m) filters on the 4th of August. Half of the approximately 1 hour observing time in each filter was on source, due to the chop-nod mode. The data were reduced using the VISIR pipeline. Raw images from the chop/nod cycle were recombined and sensitivities were estimated based on standard star observations taken on the same night.

We obtained optical and near-IR photometric observations with the dual channel imager ANDICAM \citep{DePoy03} on the SMARTS 1.3~m telescope \citep{Subasavage10}. An observing sequence consisted of a 300 s exposure in R-band, a 600 s exposure in each of V- and I-band, a 50 s exposure in each of 5 dithered images of J-band, and a 70 s exposure in each of 9 dithered images of H-band. The data were reduced with \textsc{IRAF} scripts \citep{Tody86, Tody93},\footnote{http://iraf.net/} following the standard reduction procedures in \cite{Buxton12}. The magnitudes were put to the standard scale with respect to four nearby stars in the field, with absolute calibration via optical primary standards \citep{Landolt92} on clear nights and the Two Microns All-Sky Survey catalog \citep[2MASS; ][]{Skrutskie06}. 

We are observing \srcone\ regularly with the 2~m robotic Faulkes Telescope North (located at Haleakala on Maui) and Faulkes Telescope South \citep[at Siding Spring, Australia; e.g.][]{aql:ATel1970,aql:ATel2871}, as part of an ongoing monitoring campaign of $\sim 40$ LMXBs \citep{Lewis08}. Once the 2016 outburst was first detected, we increased the cadence of observations and the number of optical filters, and also made observations with the 1~m LCO robotic telescope network. These included 1~m telescopes located at Cerro Tololo Inter-American Observatory in Chile, McDonald Observatory in Texas, USA, South African Astronomical Observatory (SAAO) at Sutherland in South Africa, and Siding Spring Observatory in Australia. During the outburst, imaging data were taken in the Bessell $B$, $V$, $R$, Sloan Digital Sky Survey (SDSS) $i\arcmin$ and Pan-STARRS $Y$-band filters (see Table \ref{tab:fluxlog} for central wavelengths).

All Faulkes/LCO data were reduced using the LCO automatic pipeline. Photometry was performed using \textsc{PHOT} in \textsc{IRAF}. For photometric calibration we made use of field stars close to \srcone\ with known magnitudes (with magnitude errors of $\leq 0.05$) tabulated in the Pan-STARRS1 and APASS catalogues \citep{PanStarrs,APASS}, \cite{aql:chevalier99aa} and \cite{aqlx1:maitra08apj}. The resulting $V$ and $i\arcmin$-band light curves are presented in Fig. \ref{fig:lc}(e). The shape of the outburst can clearly be seen and appears to correlate well with the $\swift$/XRT flux in Fig. \ref{fig:lc}(b). The optical colour is bluer when brighter (the $V$-band flux increases by a larger amplitude than the $i\arcmin$-band flux), which is consistent with previous findings \citep[e.g.][]{aqlx1:maitra08apj}. The LCO data in the remaining filters are used for the SEDs in Fig.~\ref{fig:sed} only.

To calculate the intrinsic OIR fluxes, we de-reddened the OIR magnitudes adopting an extinction of $A_{\rm V}~=~1.53\,\pm\,0.07$ mag and the extinction law of \cite{Cardelli89}. This extinction value is derived from the neutral hydrogen absorption column of \nhabs$= (4.4 \pm 0.1) \times 10^{21}$ cm$^{-2}$ measured by \cite{aql:Campana2014} using a combined fit to XMM-Newton and Chandra spectra \citep[see also Table \ref{tab:xrt} and][for similar values]{aql:King2016} and adopting the relation between $A_{\rm V}$ and \nhabs\ given in \cite{Foight2016}.

\subsection{$\swift$ and $\nustar$}
\label{swift}

$\swift$/XRT observed \srcone\ at numerous times during its 2016 outburst, four of them coincident with the days of our multi-wavelength campaign (see Table~\ref{tab:obslog}). We re-generated spectra from $\swift$/XRT observations in window timing (WT) mode following standard procedures and using the latest calibration files (March 2017; CALDB version 20170306). A type-I burst episode was recorded during observation 00033665074 and removed before spectral extraction. We next estimated the source count rate within a 30-pixel radius region (corresponding to $\sim$71\arcsec), and background counts within an annular region with inner and outer radii of 33 and 50 pixels, respectively. The first three observations (00033665074, 00033665075, 00033665076 taken on the 3rd, 5th and 7th of August, respectively) showed source count rates above the nominal pile-up threshold \citep[100 cts s$^{-1}$, see e.g.][]{romano06aa}. Thus, we next extracted source spectra from annular regions with increasing inner radii and an outer radius fixed to 30 pixels and fitted them with an absorbed power-law model. We observed significant spectral distortion as a function of the inner radius of the annular region used for spectral extraction, confirming the presence of pile-up in the three examined observations. For each observation we then excised the core region until no variation of the power-law photon index was detected. The final regions used for spectral extraction have inner radii of 2 pixels (4.72\arcsec, obs. 00033665074) and 6 pixels (14.16\arcsec, obs. 00033665075 and 00033665076). Finally, we rebinned the source spectra to ensure a minimum of 20 counts per bin and used the 0.4--9~keV energy range for analysis, performed using XSPEC \citep{arnaud96conf}, version 12.9.1. 

$\nustar$ observed Aql X-1 on the 7th of August 2016 (Obs. ID. 90201024002) for a total exposure of ~10~ks. We performed standard screening and filtering of the events using the $\nustar$ data analysis software ({\tt nustardas}) version 1.5.1. We extracted source and background events from a circular region of radius 50\arcsec\ centred on the source position and far from the source, respectively. Then, we extracted source and background spectra and generated response files for all instruments (FPMA and FPMB) using the {\tt nuproducts} pipeline. Finally, we rebinned the source spectra to ensure a minimum of 20 counts per bin and used the 3--30~keV energy range for analysis.  

We first fitted all the $\swift$/XRT observations individually.
We tried the same models for all the observations despite their potential different states for the sake of completeness. We started with one-component models like blackbody, disc blackbody, power law and thermal Comptonisation (respectively represented by the XSPEC models {\tt bbodyrad}, {\tt diskbb} \citep[][]{mitsuda84pasj}, {\tt po} and {\tt nthcomp} \citep[][]{zdziarski96mnras,zycki99mnras}). To account for absorption from neutral material we used the {\tt tbabs} XSPEC model with ISM abundances from \citet{wilms00apj} and photo-electric cross sections from \citet{verner96apj}. In Table~\ref{tab:xrt} we show the results of the fits only for those observations/models that resulted in \rchisq$<$~2 (Models~1--3, corresponding to {\tt diskbb}, {\tt po} and {\tt nthcomp}, respectively). Due to the limited statistics and energy range of the spectra, the \rchisq\ values are already very low for the one-component models shown, indicating that two-component models will be subject to a large degeneracy. However, we can attempt to constrain the two-component models by fitting all the spectra simultaneously and linking the column density of the absorbing material, expected to be constant. Following previous authors \citep{aqlx1:lin07apj,aqlx1:sakurai12pasj,aqlx1:sakurai14pasj,aqlx1:ono17pasj}, we tested four two-component models: blackbody plus power law (Model 4), disc blackbody plus power law (Model 5), blackbody plus disc blackbody (Model 6) and 
disc blackbody plus thermal Comptonisation of a blackbody (Model 7). The quality of the fits is similar for Models~4--6, confirming that the chosen models are degenerate in the available energy band. In fact, the degeneracy is such that Model~7 fails to converge due to a higher number of degrees of freedom compared to the other two-component models (and is therefore not shown). 

\begin{figure*}[ht]
\includegraphics[angle=0.0,width=0.35\textheight]{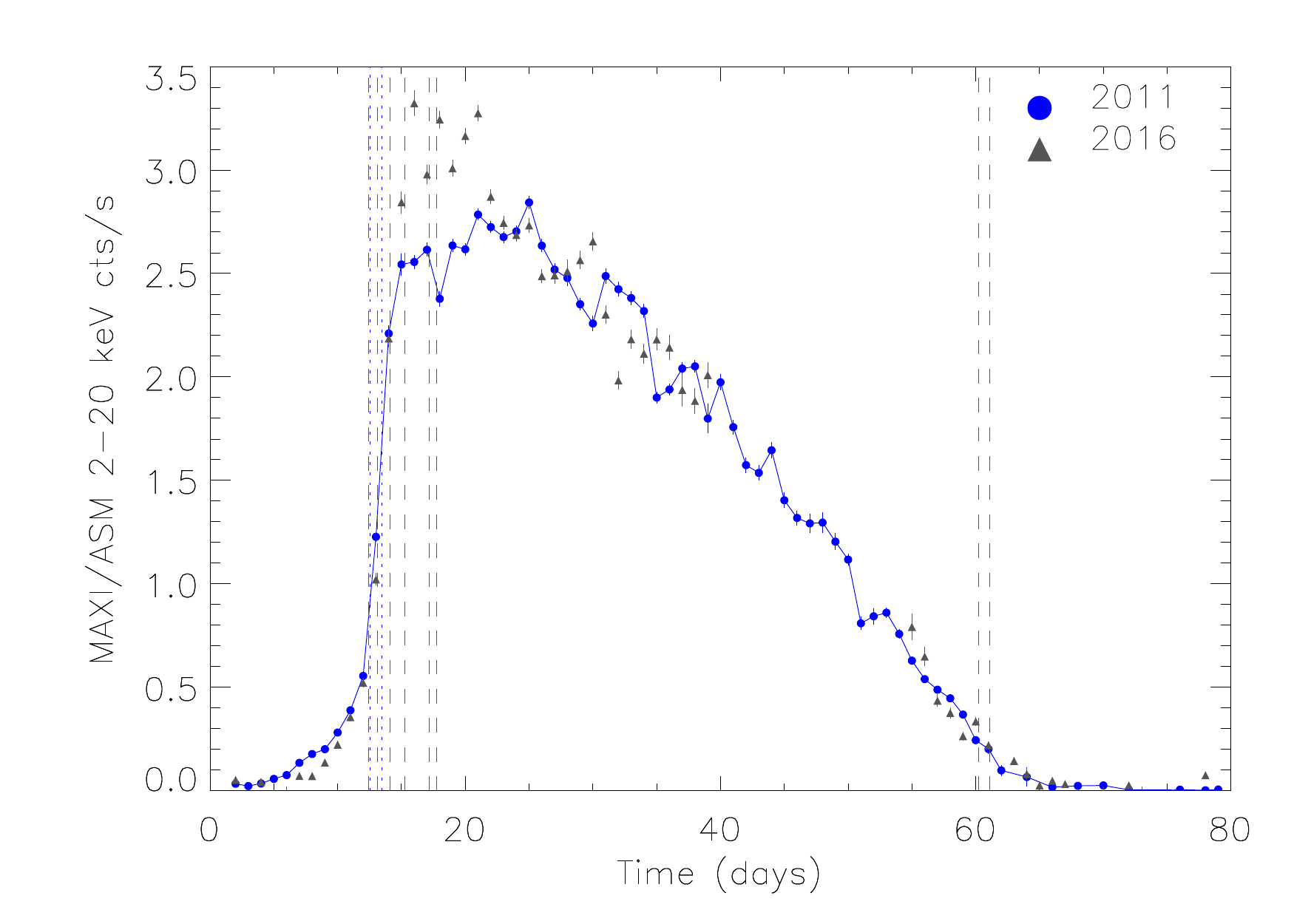}
\includegraphics[angle=0.0,width=0.35\textheight]{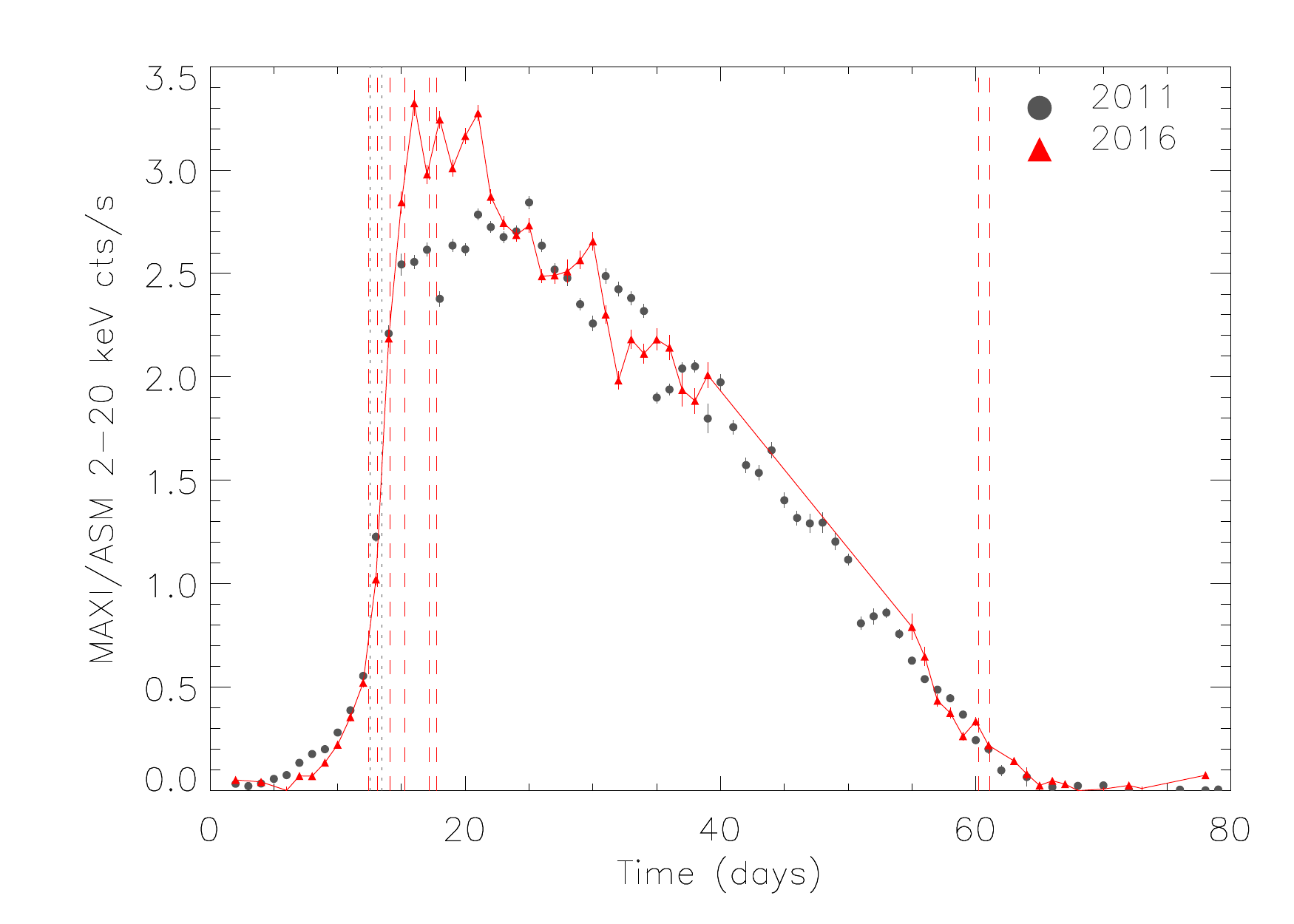}
\includegraphics[angle=0.0,width=0.35\textheight]{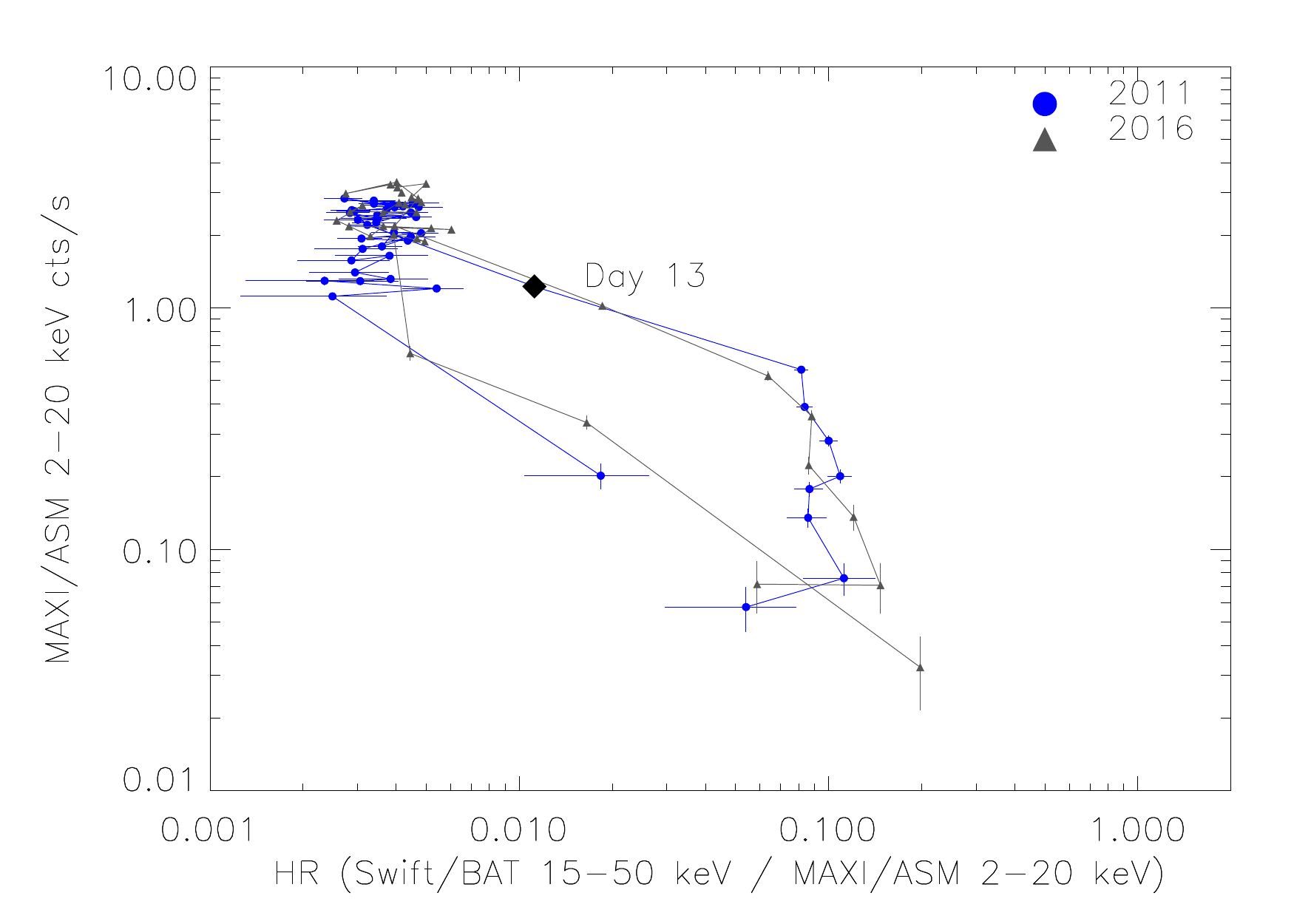}
\hspace{0.8cm}
\includegraphics[angle=0.0,width=0.35\textheight]{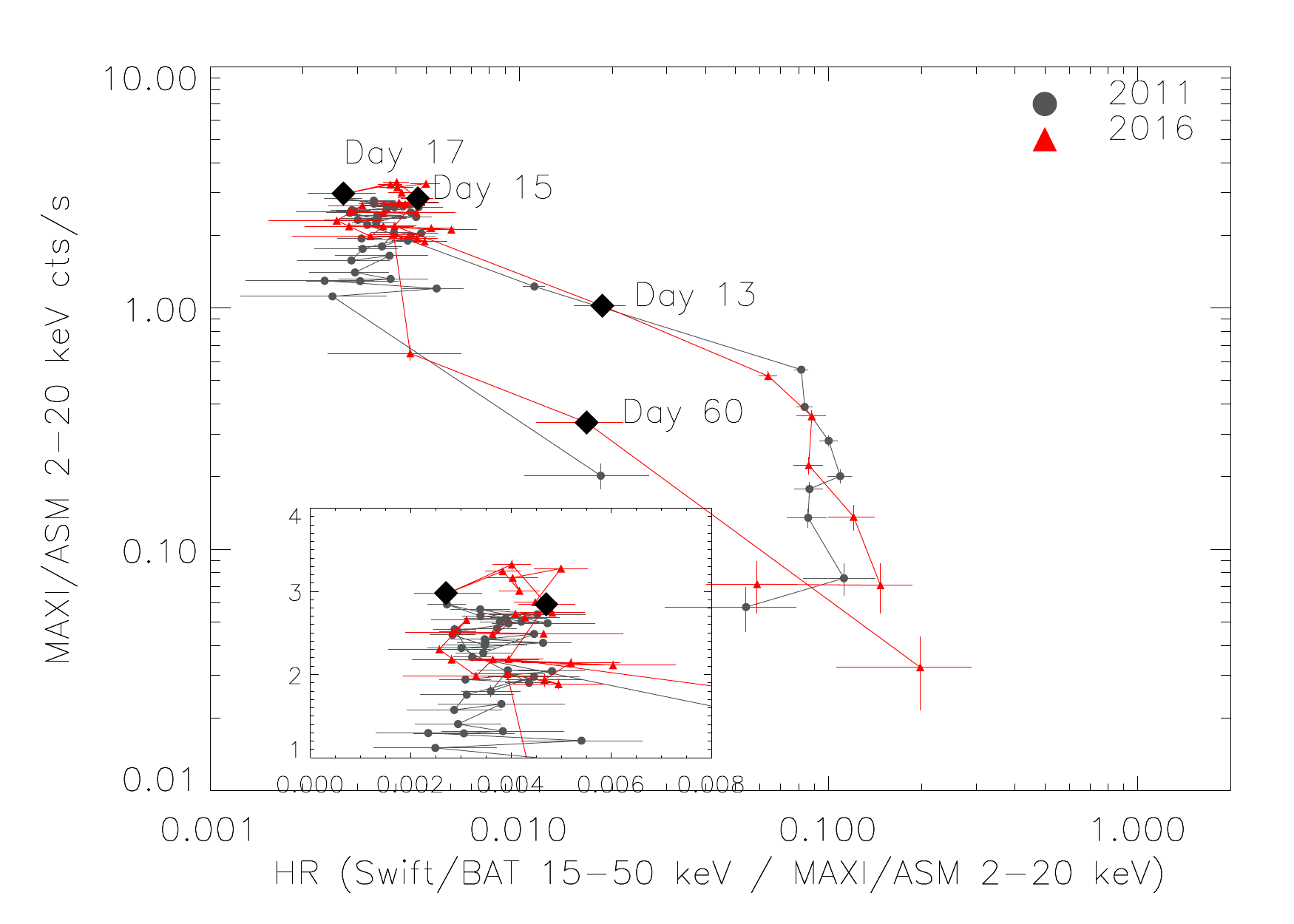}
\caption{MAXI 2--20~keV light curves (upper plots) and hardness-intensity diagrams (lower plots) for the 2011 and 2016 outbursts of \srcone. The light curves show days since the start of outburst (MJD 55843 and 57590 for the 2011 and 2016 outbursts, respectively). The vertical lines mark the time elapsed by the pointed observations: blue lines correspond to the $\suzaku$ observations in 2011 and red to the observations in 2016 presented here. The black points mark the day during which the pointed $\suzaku$ observations during the 2011 outburst (lower-left) and the $\swift$/XRT observations listed in Table~\ref{tab:obslog} (lower-right) were performed. The inset in the lower-right panel shows a zoom of the region of the HID where obs~2a and 2b are placed. In the lower panels, only points that have a hardness ratio with $>$\,2$\sigma$ significance are shown.}
\label{fig:hid}
\end{figure*}

Therefore, we next tried to further constrain the models by using hard X-ray information whenever available. For obs~2b we used the simultaneous $\swift$ and $\nustar$ observations to constrain the fit (see below). For all other observations, we compared the X-ray light curves, hardness-intensity diagrams (HIDs) and X-ray spectra of the 2016 outburst and previous outbursts with broad-band X-ray observations. 
For the observations during the hard-to-soft transition, it is remarkable that the light curves and HIDs are almost identical for the 2016 and the 2011 outbursts (see Fig.~\ref{fig:hid}), the latter of which was covered with $\suzaku$ \citep[][]{aqlx1:ono17pasj}. Indeed, looking at the outburst classification discussed in Sect.~\ref{sec:intro}, both belong to the same category of FRED and S-type or long-bright outbursts, although the 2016 outburst is brighter than the 2011 one in soft X-rays, constituting the most energetic outburst ever observed from \srcone\ \citep{aqlx1:gungor17na}.

\begin{figure*}[ht]
\includegraphics[angle=0.0,width=0.2\textheight]{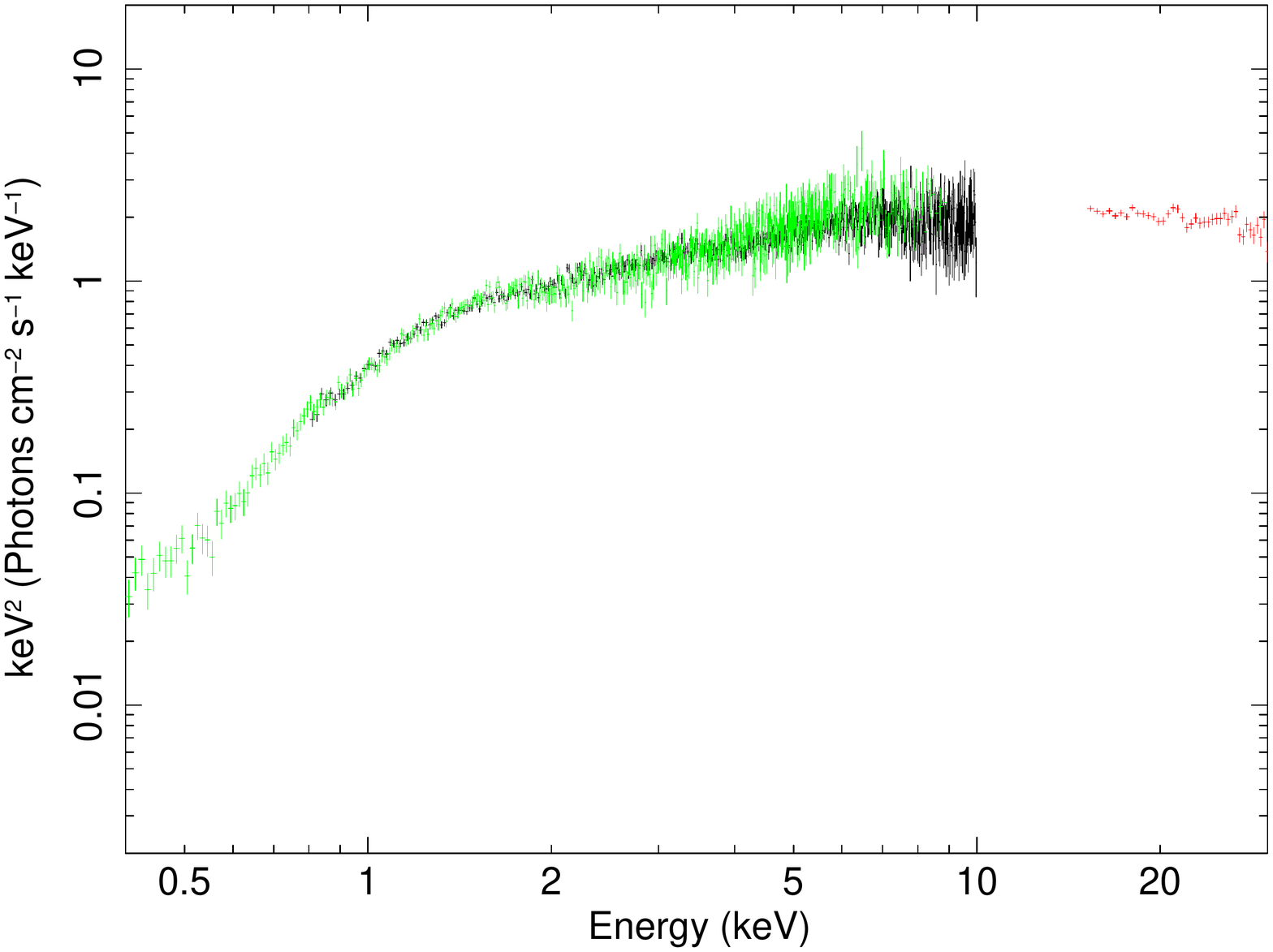}
\hspace{-0.8cm}
\includegraphics[angle=0.0,width=0.2\textheight]{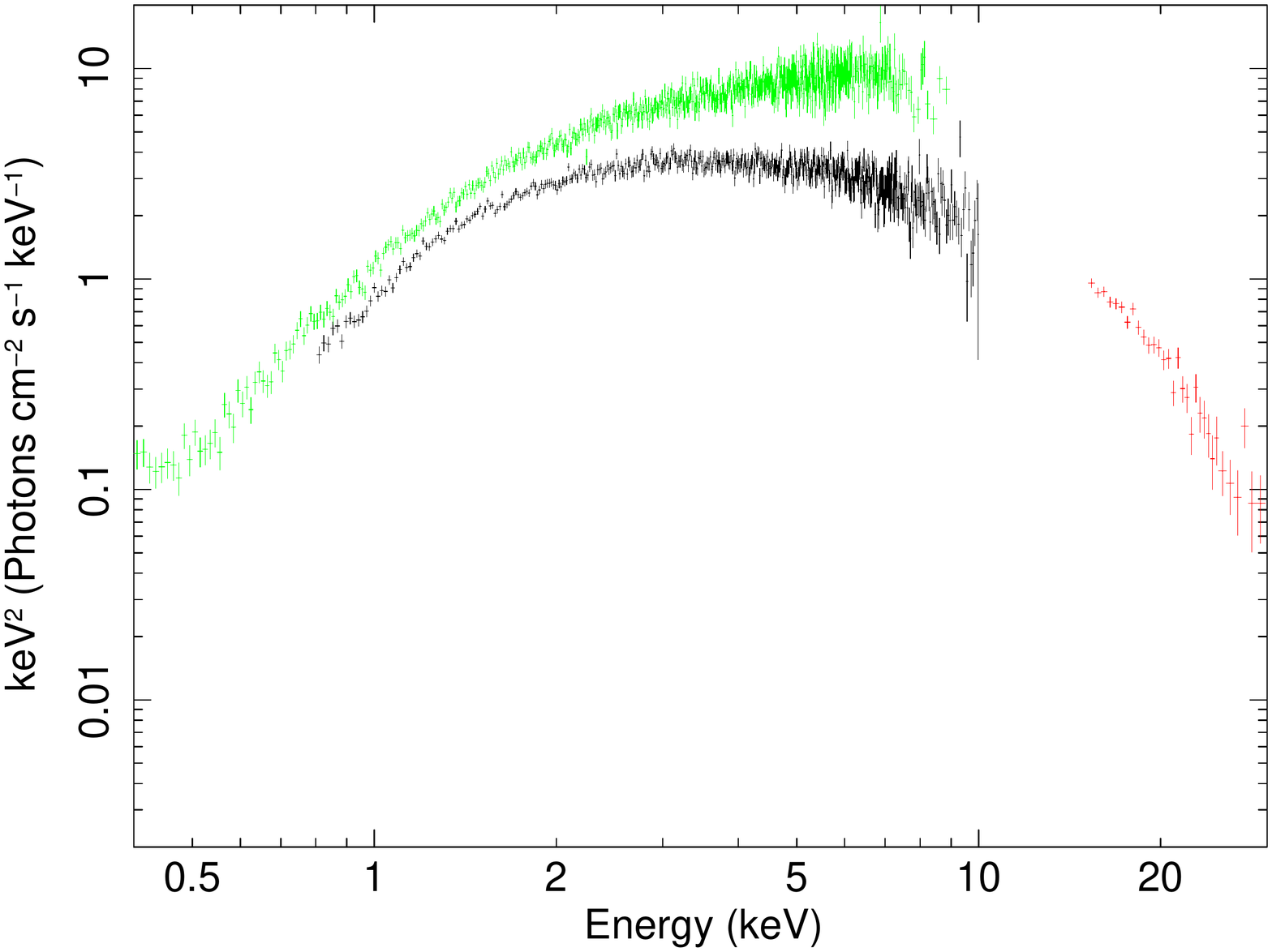}
\hspace{-0.8cm}
\includegraphics[angle=0.0,width=0.2\textheight]{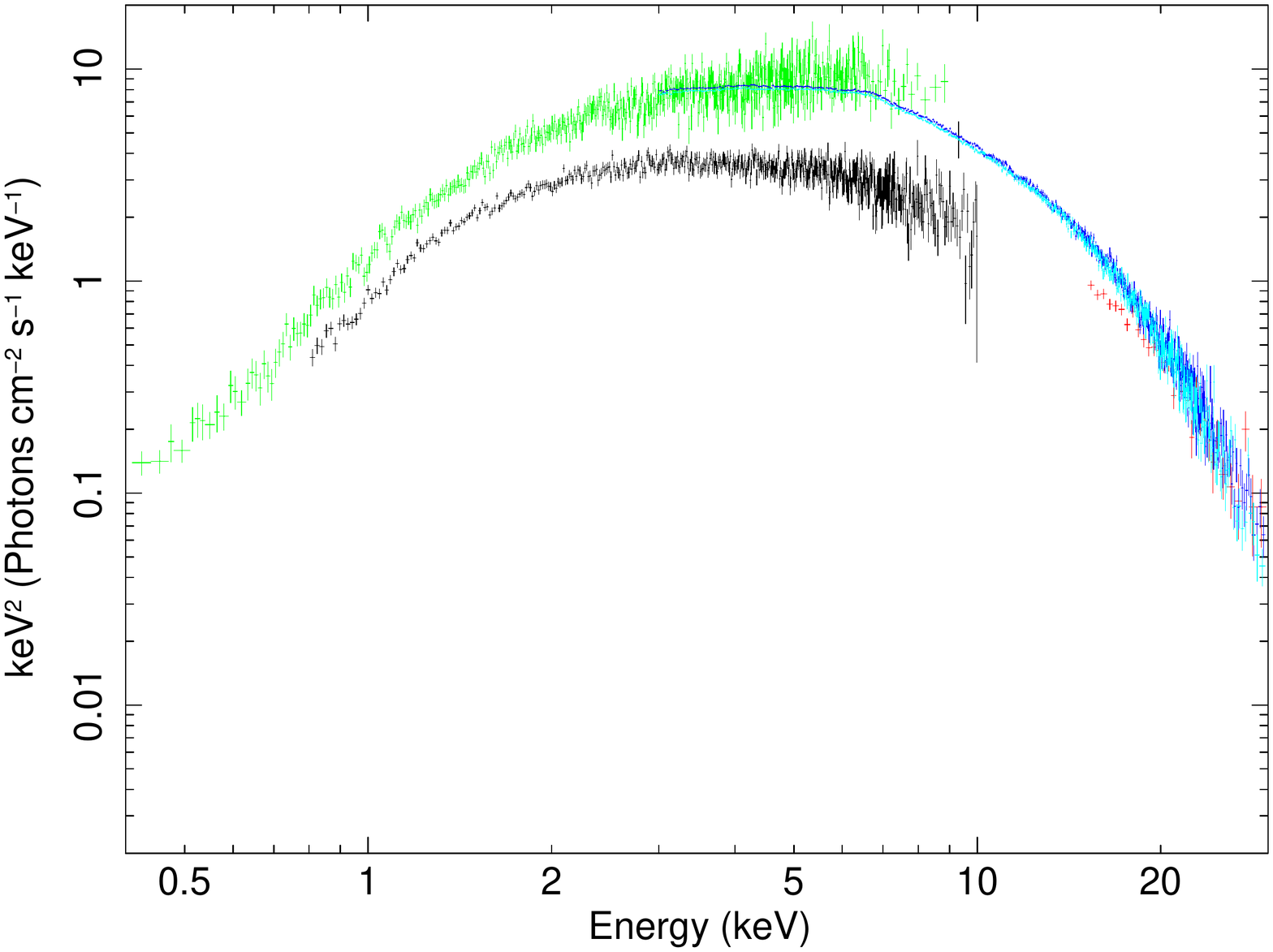}
\hspace{-0.8cm}
\includegraphics[angle=0.0,width=0.2\textheight]{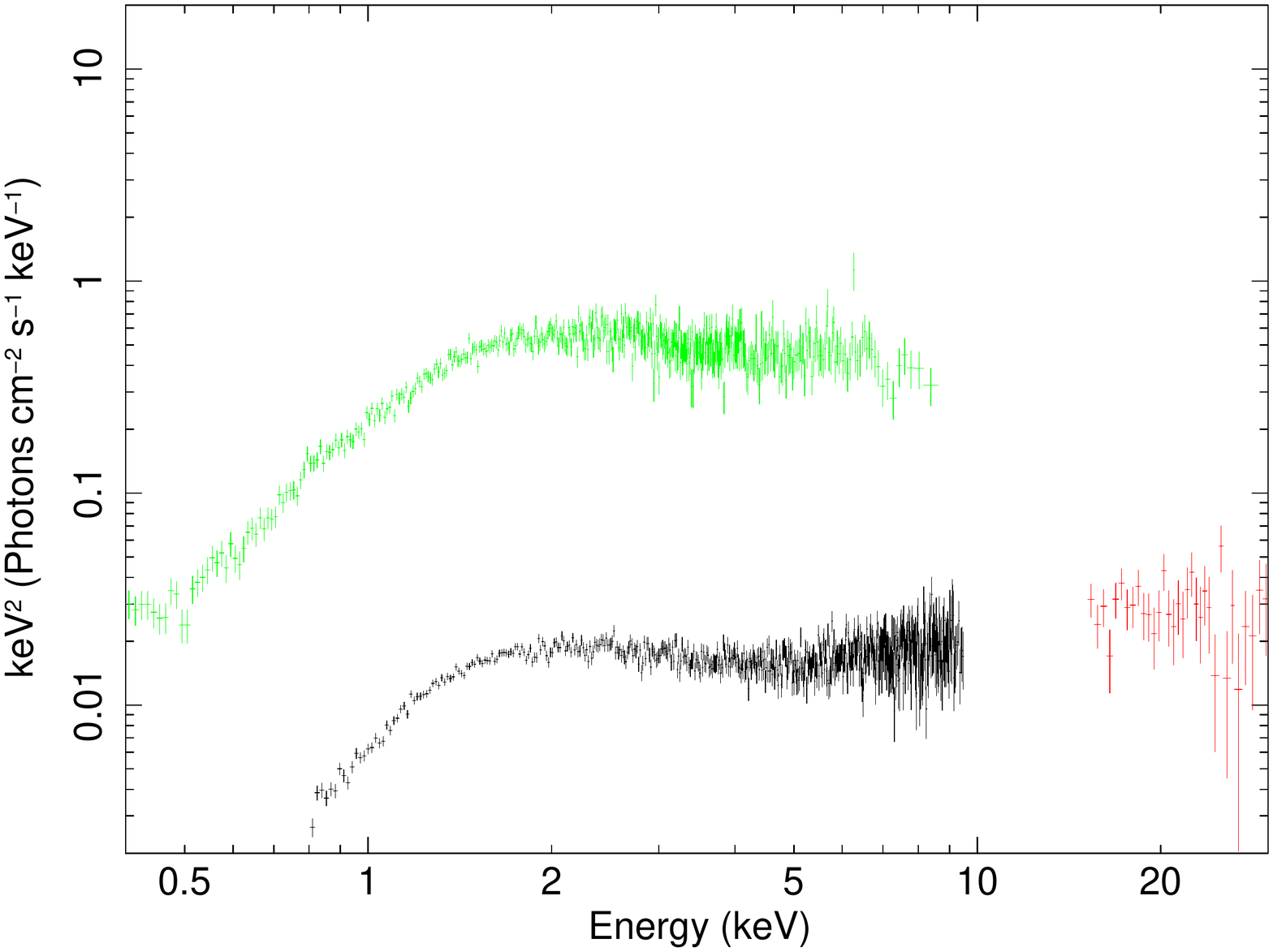}
\caption{Background-subtracted $\swift$/XRT spectra of \srcone\ for obs~1 (left panel), 2a (middle-left), 2b (middle-right panel) and 3 (right panel) after deconvolution with the response matrix using as model a power law of index 2. In each panel we show the $\swift$/XRT spectra (in green) together with $\suzaku$ broad-band spectra (in black and red) taken at similar epochs during outburst evolution. For obs~2b we also show the $\nustar$ simultaneous spectra (pink, light blue for FPMA/B, respectively). 
Based only on the comparison to broad-band spectra from previous outbursts we identify obs~1 and 3 with hard states and obs~2a and 2b with soft states (see text).}
\label{fig:xspec}
\end{figure*}

It is apparent that obs~1 occurred at an epoch similar to the $\suzaku$ observation in 2011 \citep[obs 406010020, ][]{aqlx1:ono17pasj} but ended a few hours before. Specifically, the $\suzaku$ observation spanned from day 12.54 to day 13.46 (since the start of outburst, see Fig.~\ref{fig:hid}), although the X-ray spectrum only evolved significantly after day 12.82 \citep[interval P0,][]{aqlx1:ono17pasj}. Our obs~1 spanned from day 12.44 to day 13.15 and obs~2a started at a later epoch, on day 14.10. 
Thus, we expect that the spectra of obs~1 to 2b resemble the spectra of the beginning/end of the $\suzaku$ 2011 observation, marking the start and the end of the transition from a hard to a soft state. 
To verify that this is the case we extracted the spectra from the $\suzaku$ X-ray Imaging Spectrometer \citep[XIS, ][]{Koyama2007} and the Hard X-ray Detector \citep[PIN, ][]{Takahashi2007} instruments corresponding to periods P0 and P9 from \citet{aqlx1:ono17pasj} following standard procedures. Then, we compared them to the $\swift$ and $\swift$/$\nustar$ spectra of obs~1--2b by unfolding the spectra through the response matrix using a power law of index 2 (see Fig.~\ref{fig:xspec}). Our obs~1 perfectly matches the P0 interval from \citet[][]{aqlx1:ono17pasj}, while our obs~2a and 2b are at an even softer and brighter state than the P9 interval from \citet[][]{aqlx1:ono17pasj}. 

For obs~3, occurring during the soft-to-hard transition, we extracted the $\suzaku$ spectra obtained during the decay of the 2007 outburst \citep[][Fig.~2]{aqlx1:sakurai14pasj}. Obs~3 is still about twice the luminosity of the first $\suzaku$ observations during the decay (obs 402053020, 402053030 and 402053040) but the spectral shape looks similar to an even later stage during the decay (obs 402053050, or $\suzaku$ obs~5). Therefore, in Fig.~\ref{fig:xspec} we plot obs~3 together with the latter observation from \citet{aqlx1:sakurai14pasj}. 

Taking into account the similarities between $\swift$/XRT spectra and previous broad-band spectra highlighted above we next constrained the fits as follows: we used a basic model consisting of thermal Comptonisation absorbed from neutral material ({\tt tbabs*nthcomp} in XSPEC) and, where necessary, we added a disc blackbody ({\tt tbabs*(diskbb+nthcomp)}). For obs~2b, to account for the strong iron K line present in the $\nustar$ spectra we also included the reflection component {\tt rfxconv} \citep{gx339:kolehmainen11mnras} and the final model was {\tt tbabs*(diskbb+rfxconv*nthcomp)}. Then, we fitted obs~1 and $\suzaku$ P0 tying all the parameters of the fit. Next, we fitted obs~2b with the $\nustar$ spectra again tying all the parameters of the fit. Finally, we fitted $\suzaku$ P9 and $\suzaku$ obs~5 individually. Since our obs~2a and 3 are not similar enough to or have simultaneous hard X-ray observations, we imposed values for the parameters intermediate between those of $\suzaku$ P9 and obs~2b and between obs~2b and $\suzaku$ obs~5, respectively. 
Table~\ref{tab:xrt2} shows the results of the fits.   

\begin{table*}
\begin{center}
\caption[]{Best fits to the 0.4--9~keV $\swift$/XRT spectra for observations 1--3 using one- and two-component models. Only fits with \rchisq\ $<$ 2 are shown. \nhabs\ is the column density for the neutral absorber in units of $10^{22}$ cm$^{-2}$. \kbb, \ktdbb\ and $k_{\rm po}$ are the normalisations of the blackbody, disc blackbody and the power-law components in units of 
[(R$_{in}$/D$_{10}$)$^{2}$], [(R$_{in}$/D$_{10}$)$^{2}$ cos$\theta$] and [ph keV$^{-1}$cm$^{-2}s^{-1}$ at 1 keV], respectively, R$_{in}$ being the radius in km, D$_{10}$ the distance in units of 10 kpc and $\theta$ the inclination of the source. $k_{\rm nth}$ is the normalisation of the {\tt nthcomp} component. 
  \ktbb/\ktdbb/$kT_{\rm e}$ are the temperatures of the blackbody/disc blackbody components and electrons in units of keV and $\Gamma$ the index of the power law. $F$ is the unabsorbed
  flux of the continuum in units of erg cm$^{-2}$ s$^{-1}$. {\it t} indicates that a parameter was tied for all the observations. {\it f} indicates that a parameter was fixed during the fits.}
\small\addtolength{\tabcolsep}{-3pt}
\begin{tabular}{lcccccccclll}
\hline \hline\noalign{\smallskip}
 Obs. & Obs. ID & \multicolumn{5}{c}{Best-fit model} & $F_{0.4-9 keV}$ & $F_{15--50 keV}$ & \rchisq\ (d.o.f.) \\
\hline\noalign{\smallskip}
& & \multicolumn{5}{c}{\tt Model 1: tbabs*(diskbb)} & \\
 \noalign{\smallskip\hrule\smallskip}
& & \nhabs\ & \ktbb\ & \kbb\ &  &  & \\
 \noalign{\smallskip\smallskip}
2a & 00033665075 & 0.41\,$\pm$\,0.01 & 2.31\,$\pm$\,0.04 & 47\,$\pm$\,3 & & & 2.42\,$\times$\,10$^{-8}$ & 6.48\,$\times$\,10$^{-10}$ & 0.95 (662) \\
2b & 00033665076 &  0.44\,$\pm$\,0.01 & 2.01\,$\pm$\,0.04 & 83\,$\pm$\,6 & & & 2.54\,$\times$\,10$^{-8}$ & 3.02\,$\times$\,10$^{-10}$ & 1.13 (568)\\
\noalign{\smallskip\hrule\smallskip}
& & \multicolumn{5}{c}{\tt Model 2: tbabs*(po)} & \\
\noalign{\smallskip\hrule\smallskip}
& & \nhabs\ & $\Gamma$ & $k_{\rm po}$ & \\
 \noalign{\smallskip\smallskip}
1 & 00033665074 & 0.44\,$\pm$\,0.01 & 1.51\,$\pm$\,0.02 & 0.79\,$\pm$\,0.02 & & & 5.89\,$\times$\,10$^{-9}$ & 7.72\,$\times$\,10$^{-9}$ & 1.15 (616) \\
2a & 00033665075 & 0.66\,$\pm$\,0.02 & 1.42\,$\pm$\,0.02 & 3.49\,$\pm$\,0.08 & & & 2.68\,$\times$\,10$^{-8}$ & 4.68\,$\times$\,10$^{-8}$ & 1.63 (662)  \\
2b & 00033665076 & 0.76\,$\pm$\,0.03 & 1.58\,$\pm$\,0.02 & 4.57\,$\pm$\,0.15 & & & 3.21\,$\times$\,10$^{-8}$ & 3.60\,$\times$\,10$^{-8}$ & 1.45 (568) \\
3 & 00033665089 & 0.69\,$\pm$\,0.02 & 2.28\,$\pm$\,0.03 & 0.76\,$\pm$\,0.02 & & & 3.25\,$\times$\,10$^{-9}$ & 5.77\,$\times$\,10$^{-10}$ & 1.71 (462)\\
\noalign{\smallskip\hrule\smallskip}
& & \multicolumn{5}{c}{\tt Model 3: tbabs*(nthcomp(bb))} & \\
\noalign{\smallskip\hrule\smallskip}
& & \nhabs\ & $\Gamma$ & \ktbb\ & $kT_{\rm e}$ & $k_{\rm nth}$  & \\
 \noalign{\smallskip\smallskip}
1 & 00033665074 & 0.36\,$\pm$\,0.03 & 1.58\,$\pm$\,0.03 & 0.20\,$\pm$\,0.03 & 5\,$^{+2}_{-1}$ & 0.70\,$\pm$\,0.04 & 5.60\,$\times$\,10$^{-9}$ & 2.90\,$\times$\,10$^{-9}$ & 1.08 (614) \\
 2a & 00033665075 & 0.27\,$\pm$\,0.02 & 1.59\,$\pm$\,0.05 & 0.46\,$\pm$\,0.03 & 2.03\,$^{+0.18}_{-0.14}$ & 1.71\,$\pm$\,0.08 & 2.31\,$\times$\,10$^{-8}$ & 6.89\,$\times$\,10$^{-10}$ & 0.92 (660)  \\
 2b & 00033665076 & 0.64\,$\pm$\,0.03 & 1.55\,$\pm$\,0.02 & $<$\,0.09 & 1.68\,$\pm$\,0.08 & 3.8\,$\pm$\,0.2 & 2.87\,$\times$\,10$^{-8}$ & 2.44\,$\times$\,10$^{-10}$ & 1.17 (566)   \\
 3 & 00033665089 & 0.27\,$\pm$\,0.02 & 2.51\,$^{+0.03}_{-0.06}$ & 0.36\,$\pm$\,0.01 & $>$\,11 & 0.35\,$\pm$\,0.01 & 1.94\,$\times$\,10$^{-9}$ & 3.18\,$\times$\,10$^{-10}$ & 1.25 (460)  \\
\noalign{\smallskip\hrule\smallskip}
& & \multicolumn{5}{c}{\tt Model 4: tbabs*(bbodyrad+po)} & \\
\noalign{\smallskip\hrule\smallskip}
& & \nhabs\ &  \ktbb\ & \kbb\ & $\Gamma$ & $k_{\rm po}$ \\
 \noalign{\smallskip\smallskip}
1 & 00033665074 & 0.44\,$\pm$\,0.02 & 0.31\,$\pm$\,0.02 & 5836\,$^{+1730}_{-1350}$ & 1.29\,$\pm$\,0.05 & 0.57\,$\pm$\,0.04 & 5.96\,$\times$\,10$^{-9}$ & 1.19\,$\times$\,10$^{-8}$ & 1.04 (2303) \\
2a & 00033665075 & $t$ & 0.95\,$\pm$\,0.04 & 795\,$^{+120}_{-100}$ & 1.28\,$\pm$\,0.05 & 1.87\,$\pm$\,0.11 & 2.50\,$\times$\,10$^{-8}$ & 4.08\,$\times$\,10$^{-8}$ &  \\
2b & 00033665076 & $t$ & 0.79\,$\pm$\,0.03 & 1842\,$^{+290}_{-260}$ & 1.22\,$\pm$\,0.05 & 1.82\,$\pm$\,0.15 & 2.64\,$\times$\,10$^{-8}$ & 4.82\,$\times$\,10$^{-8}$ &   \\
3 & 00033665089 & $t$ & 0.49\,$\pm$\,0.01 & 1029\,$^{+190}_{-170}$ & 1.81\,$\pm$\,0.05 & 0.30\,$\pm$\,0.03 & 2.33\,$\times$\,10$^{-9}$ & 1.09\,$\times$\,10$^{-9}$ &   \\
\noalign{\smallskip\hrule\smallskip}
& & \multicolumn{5}{c}{\tt Model 5: tbabs*(diskbb+po)} & \\
\noalign{\smallskip\hrule\smallskip}
& & \nhabs\ &  \ktdbb\ & \kdbb\ & $\Gamma$ & $k_{\rm po}$ \\
 \noalign{\smallskip\smallskip}
1 & 00033665074 & 0.45\,$\pm$\,0.01 & 0.46\,$\pm$\,0.04 & 1059\,$^{+360}_{-260}$ & 1.23\,$\pm$\,0.06 & 0.52\,$\pm$\,0.05 & 6.09\,$\times$\,10$^{-9}$ & 1.31\,$\times$\,10$^{-8}$ & 1.05 (2303) \\
2a & 00033665075 & $t$ & 2.3\,$\pm$\,0.2 & 41\,$^{+12}_{-4}$ & 1.5\,$^{+1.5}_{-0.4}$ & 0.44\,$^{+0.19}_{-0.25}$ & 2.48\,$\times$\,10$^{-8}$ & 4.95\,$\times$\,10$^{-9}$ &  \\
2b & 00033665076 & $t$ & 1.66\,$^{+0.19}_{-0.09}$& 139\,$\pm$\,20 & 0.3\,$^{+0.4}_{-2.1}$ & 0.15\,$^{+0.26}_{-0.14}$ & 2.62\,$\times$\,10$^{-8}$ & 8.36\,$\times$\,10$^{-8}$ &   \\
3 & 00033665089 & $t$ & 0.75\,$\pm$\,0.03 & 183\,$\pm$\,30 & 1.58\,$^{+0.11}_{-0.14}$ & 0.18\,$\pm$\,0.04 & 2.34\,$\times$\,10$^{-9}$ & 1.43\,$\times$\,10$^{-9}$ &   \\
\noalign{\smallskip\hrule\smallskip}
& & \multicolumn{5}{c}{\tt Model 6: tbabs*(diskbb+bbodyrad)} & \\
\noalign{\smallskip\hrule\smallskip}
& & \nhabs\ &  \ktdbb\ & \kdbb\ & \ktbb\ & \kbb\ \\
 \noalign{\smallskip\smallskip}
1 & 00033665074 & 0.429\,$\pm$\,0.009 & 0.62\,$\pm$\,0.02 & 848\,$^{+125}_{-110}$ & 1.83\,$\pm$\,0.07 & 40\,$\pm$\,5 & 5.84\,$\times$\,10$^{-9}$ & 1.67\,$\times$\,10$^{-10}$ & 1.04 (2304) \\
2a & 00033665075 & $t$ & 1.4\,$^{+0.5}_{-0.2}$ & 203\,$^{+140}_{-120}$ & 1.9\,$^{+0.9}_{-0.2}$ & 96\,$\pm$\,82 & 2.45\,$\times$\,10$^{-8}$ & 7.00\,$\times$\,10$^{-10}$ &  \\
2b & 00033665076 & $t$ & 1.44\,$^{+0.12}_{-0.09}$& 206\,$\pm$\,40 & 2 ($f$) & 68\,$^{+11}_{-14}$ & 2.57\,$\times$\,10$^{-8}$ & 6.57\,$\times$\,10$^{-10}$ &   \\
3 & 00033665089 & $t$ & 0.74\,$\pm$\,0.02 & 280\,$^{+40}_{-30}$ & 1.80\,$^{+0.18}_{-0.15}$ & 8\,$\pm$\,3 & 2.25\,$\times$\,10$^{-9}$ & 2.77\,$\times$\,10$^{-11}$ &   \\
\hline\noalign{\smallskip}
\label{tab:xrt}
\end{tabular}
\end{center} 
\end{table*}

\begin{table*}
\begin{center}
\caption[]{Best fits to the 0.4--9~keV $\swift$/XRT spectra for observations 1--3 using model {\tt tbabs*nthcomp} for obs~1 and 3, model {\tt tbabs*(diskbb+nthcomp)} for obs~2a and model {\tt tbabs*(diskbb+rfxconv*nthcomp)} for obs~2b. The model parameters are the same as in Table~\ref{tab:xrt}. In addition, rel$\_$refl is the reflection fraction and log($\xi$) the ionisation parameter of the reflection component, for which we have fixed the inclination to 40\deg.
 {\it p} indicates that a parameter pegged at its lower or upper limit. {\it f} indicates that a parameter was fixed during the fits. For obs~2a and 3 we imposed values for the parameters of the fit between those of adjacent broad-band observations, except for the normalisations of the individual components in obs~3 that we leave free (see text). }
\small\addtolength{\tabcolsep}{-3pt}
\begin{tabular}{lccccccc}
\hline \hline\noalign{\smallskip}
 Obs. & & 1 \& $\suzaku$ P0 & $\suzaku$ P9 & 2a & 2b \& $\nustar$ & 3 & $\suzaku$ 5 \\
\hline\noalign{\smallskip}

\noalign{\smallskip\hrule\smallskip}
 \noalign{\smallskip\smallskip}
& Comp. & \\
Parameter & \\ 
& {\tt tbabs} & \\
 \nhabs\ & & 0.43\,$\pm$\,0.03 & 0.43 ($f$) & 0.40\,$^{+0.01}_{p}$ & 0.40\,$\pm$\,0.02 & 0.40\,$^{+0.01}_{p}$  &0.43 ($f$) \\
& {\tt diskbb} \\
 \ktdbb\ & & -- & 1.44\,$^{+0.16}_{-0.13}$ & 2.11\,$^{p}_{-0.01}$ & 2.11\,$\pm$\,0.03 & 0.57\,$^{+0.01}_{p}$ & 0.57\,$\pm$\,0.02 \\
 \kdbb\ & & -- & 68\,$^{+31}_{-21}$ & 57.1\,$\pm$\,0.5 & 29\,$\pm$\,2 & 783\,$^{+14}_{-22}$ & 24\,$\pm$\,5 \\
& {\tt rfxconv} & & & \\
rel$\_$refl & & -- & -- & -- &0.77\,$\pm$\,0.12 & -- & -- \\
xi & & -- & -- & -- & 3.84\,$\pm$\,0.06 & -- & -- \\
& {\tt nthcomp} & \\
 $\Gamma$ & & 1.64\,$\pm$\,0.01 & 2.0\,$\pm$\,0.1 & 2.00\,$^{+0.02}_{p}$ & 2.31\,$\pm$\,0.05 & 1.78\,$\pm$\,0.06 & 1.71\,$\pm$\,0.06 \\
$kT_{e}$ & & 7.2\,$\pm$\,0.2 & 2.8\,$\pm$\,0.2 & 3.04\,$^{p}_{-0.19}$ & 3.04\,$^{+0.10}_{-0.07}$ & 3.0\,$^{+0.5}_{p}$ & 10\,$^{+14}_{-3}$ \\
\ktbb\ & & 0.16\,$\pm$\,0.03 & 0.21\,$\pm$\,0.04 & 0.21\,$^{+0.01}_{p}$ & 0.56\,$\pm$\,0.02 & 0.66\,$^{p}_{-0.02}$ & 0.35\,$^{+0.08}_{-0.06}$ \\
$k_{nth}$ & & 0.80\,$\pm$\,0.04 & 1.51\,$^{+0.08}_{-0.10}$ & 0.77\,$^{+0.03}_{p}$ & 0.77\,$\pm$\,0.05 & 0.056\,$\pm$\,0.004 & 0.006\,$\pm$\,0.001 \\
& \\
$F_{0.4-9 keV}$  & & 5.65\,$\times$\,10$^{-9}$ & 1.3\,$\times$\,10$^{-8}$ & 2.5\,$\times$\,10$^{-8}$ & 2.4\,$\times$\,10$^{-8}$ & 2.5\,$\times$\,10$^{-9}$ & 9.7\,$\times$\,10$^{-11}$\\
$F_{15-50 keV}$ & & 3.15\,$\times$\,10$^{-9}$ & 4.3\,$\times$\,10$^{-10}$ & 5.1\,$\times$\,10$^{-10}$ & 6.0\,$\times$\,10$^{-10}$ & 1.3\,$\times$\,10$^{-10}$ &5.0\,$\times$\,10$^{-11}$ \\
\rchisq\ (d.o.f.) & & 1.30 (1348) & 1.10 (654) & 1.27 (658) & 1.21 (1658) & 1.43 (458) & 1.13 (672) \\
\noalign{\smallskip\hrule\smallskip}
\hline\noalign{\smallskip}
\label{tab:xrt2}
\end{tabular}
\end{center} 
\end{table*}

\section{Results}
\label{sec:results}

\subsection{Accretion state}
\label{subsec:as}

The aim of our multi-wavelength observations was to catch the source at state transitions, which is particularly challenging given that they typically only last between 1 and 4 days 
\citep[see Table 1,][]{munoz-darias14mnras}. 

Based on the comparison of the $\swift$/XRT spectra with previous outbursts and on the fits in Section~\ref{swift} we conclude that obs~1 is a hard state: it can be well matched to $\suzaku$ P0, classified as a hard state, and it can be fitted with only a Comptonisation component with relatively hard photon index (see Table~\ref{tab:xrt2}). Instead, obs~2a is softer than $\suzaku$ P9, classified as a soft state, and we need to include a second component in the fit consisting of a relatively hot disc (2.11~keV) besides the Comptonisation component that has a softer photon index (2.00) and a lower electron temperature compared to obs~1. Thus, we conclude that obs~2a is a soft state. Obs~2b is very similar to obs~2a and is placed in the same part of the HID diagram (see Fig.~\ref{fig:hid}). Moreover, the broad-band fits including the simultaneous $\nustar$ data show again a hot disc and a Comptonisation component with a soft photon index (2.31) and a low electron temperature. Thus, we conclude that obs~2b occurs during a soft state too. 
We also note that the Eddington fraction during obs~2a/2b is $\sim$60--65\% for a 1.4~M$_{\odot}$ neutron star, compared to 30\% at the end of the $\suzaku$ observation in 2011.
While the 2--20~keV light curves seem to indicate that the peak of the outburst occurred between obs~2a and 2b, the 15--50~keV flux already peaks during obs~1 and drops thereafter, causing a significant drop in the $\swift$/(BAT/XRT) HR  by the time of obs~2a (see Fig.~\ref{fig:lc}, panel d). In contrast, the OIR flux peaks a few days after obs~2b (see Fig.~\ref{fig:lc}, panel e). 

The reverse transition, from a soft to a hard state, occurs generally when the source bolometric luminosity has decreased well below 10\% Eddington and is marked by a sudden decrease of the soft X-rays, which is also associated with a change in the X-ray HR. In the 2016 outburst, we detected an increase in the 15--50~keV/0.3--10~keV HR around day~57, as the bolometric luminosity was $\sim$5\%. Obs~3 was then performed at days~60--61 and it clearly lies in a relatively hard region of the HID. The fits from previous section indicate that the temperature of the disc has significantly dropped to 0.57~keV while the index of the Comptonisation component has become harder again (1.78). Thus, we classify obs~3 as a hard state. 

Besides the spectral degeneracy, a caveat to the classification above is that the multi-wavelength observations span several hours. For example, obs~1/3 last in total 17/19 hours, respectively, i.e. about twice the duration of the hard-to-soft transition observed by \citet[][]{aqlx1:ono17pasj}. This is not a concern for obs~2a/2b, which we can safely classify as soft states based on the high luminosity, the position in the HID and the curvature of the $\swift$/XRT-$\nustar$ spectra. In contrast, obs~1/3 show $\swift$/XRT spectra consistent with a hard state (see above) based on a comparison with previous broad-band X-ray spectra but the luminosity and the position in the HID may indicate that the transition is actually underway. The $\swift$/BAT to $\swift$/XRT HR has already started to decrease before the onset of obs~1 (see Fig.~\ref{fig:lc}, d). The $\swift$/BAT to MAXI HID shows that the daily HR is still on the hard region of the HID on day 12, at the time of the ATCA observation, but has significantly moved towards the soft region on day 13 (see Fig.~\ref{fig:hid}, lower-right). Hence, ALMA observations, occurring on day 13, 1.75 hours before the $\swift$/XRT observation, could already track the start of the transition.
Similarly, the $\swift$/BAT to $\swift$/XRT HR shows a peak during obs~3 (see Fig.~\ref{fig:lc}, d) but decreases right after (although the errors are large). The $\swift$/BAT to MAXI HID shows that the daily HR is moving towards the hard region of the HID on day 60 and has reached this region by the next day (see Fig.~\ref{fig:hid}, lower-right). Specifically, the ATCA observation could track a state softer than the $\swift$/XRT observation, starting 3.2 hours later during day 60. The ALMA observations, starting 7.7 hours after the end of the $\swift$/XRT observation, should track a slightly harder state than the latter observation. 

Thus, we conclude that obs~1/3 may be tracking a relatively hard state but during the state transition and classify them as hard/intermediate states. Table~\ref{tab:indices} summarises the state classification of our multi-wavelength pointed observations taking into account these considerations.

\subsection{Broad-band spectra}
\label{sed}

\begin{figure*}[ht]
\includegraphics[angle=0.0,width=0.35\textheight]{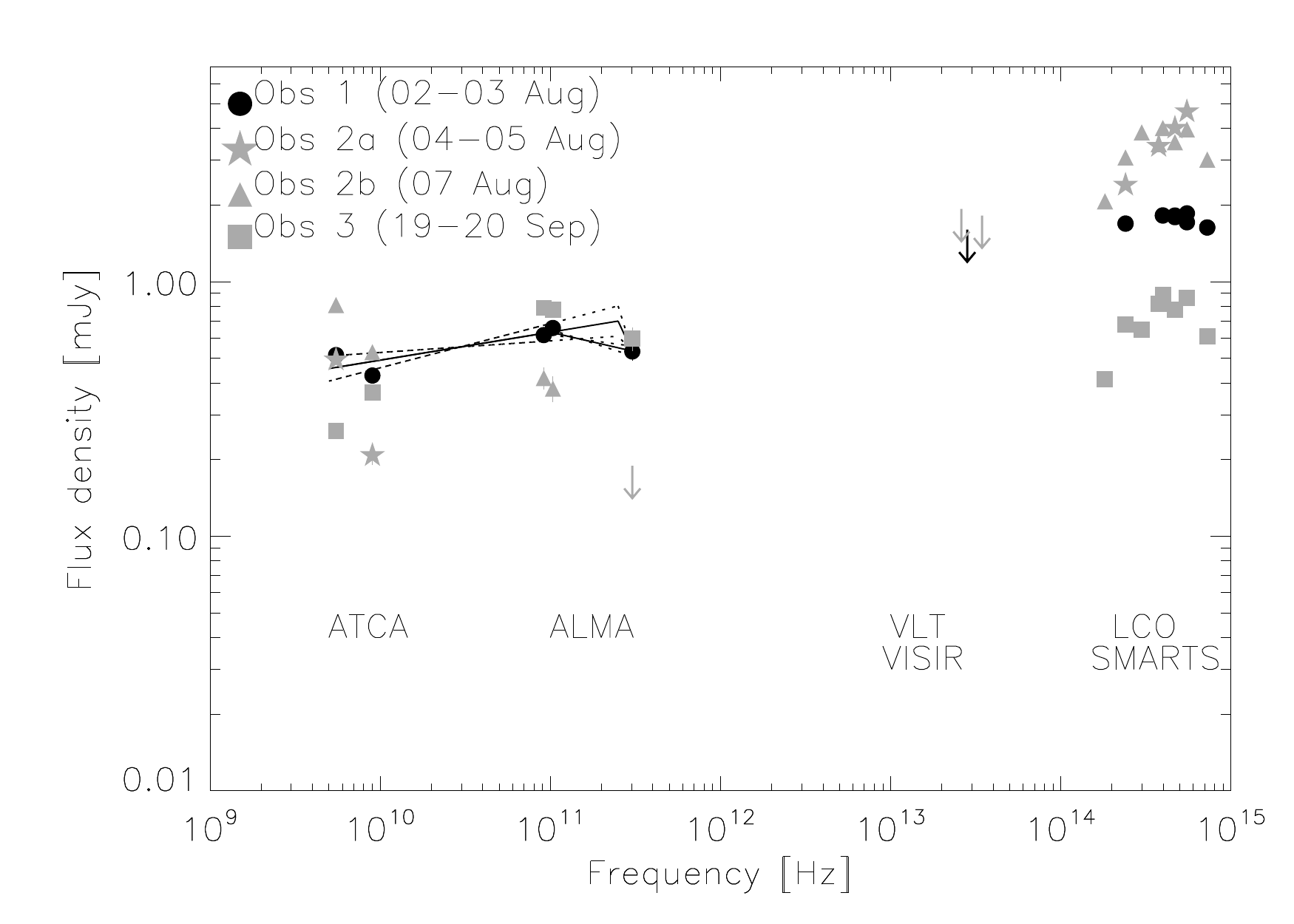}
\includegraphics[angle=0.0,width=0.35\textheight]{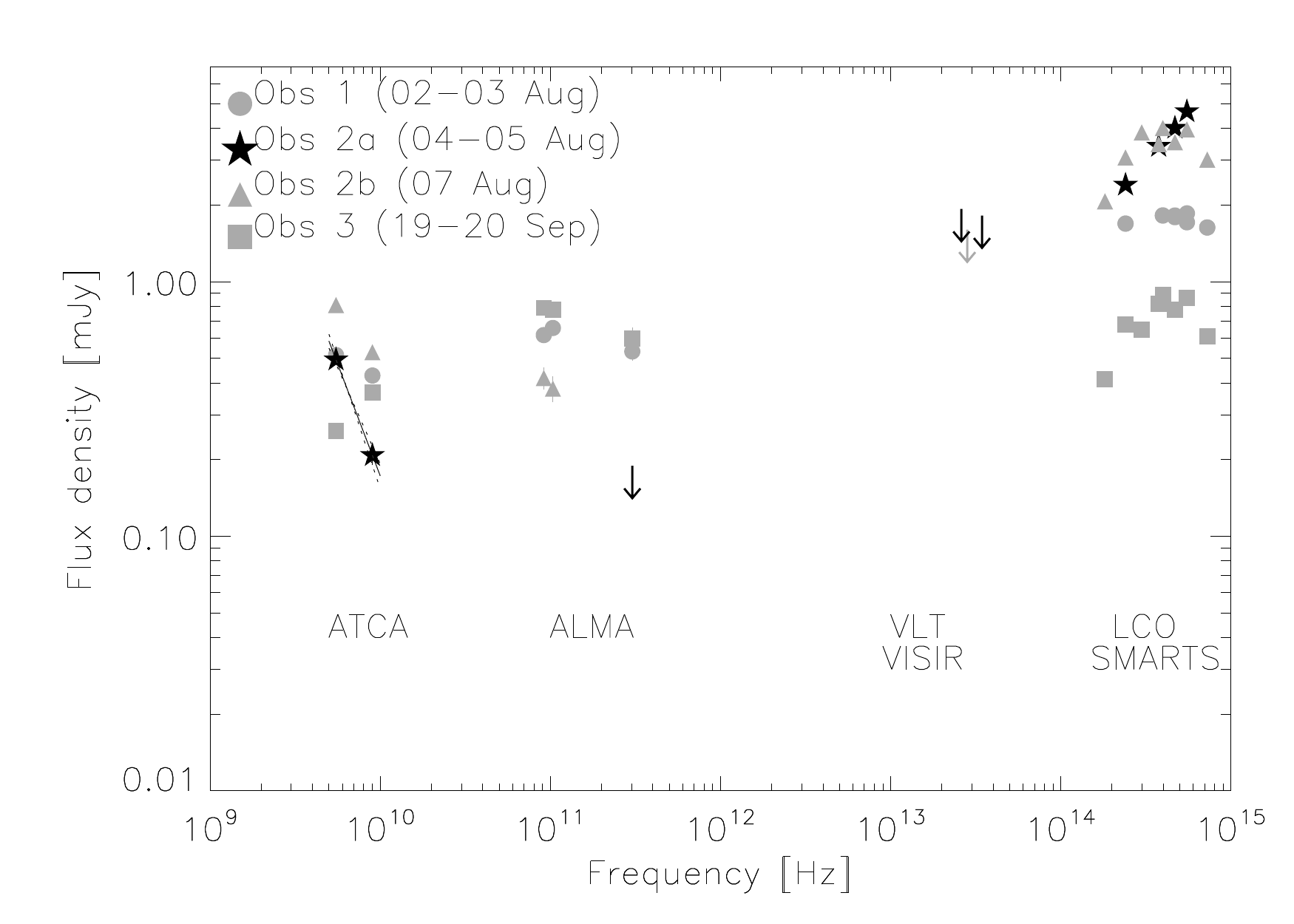}
\includegraphics[angle=0.0,width=0.35\textheight]{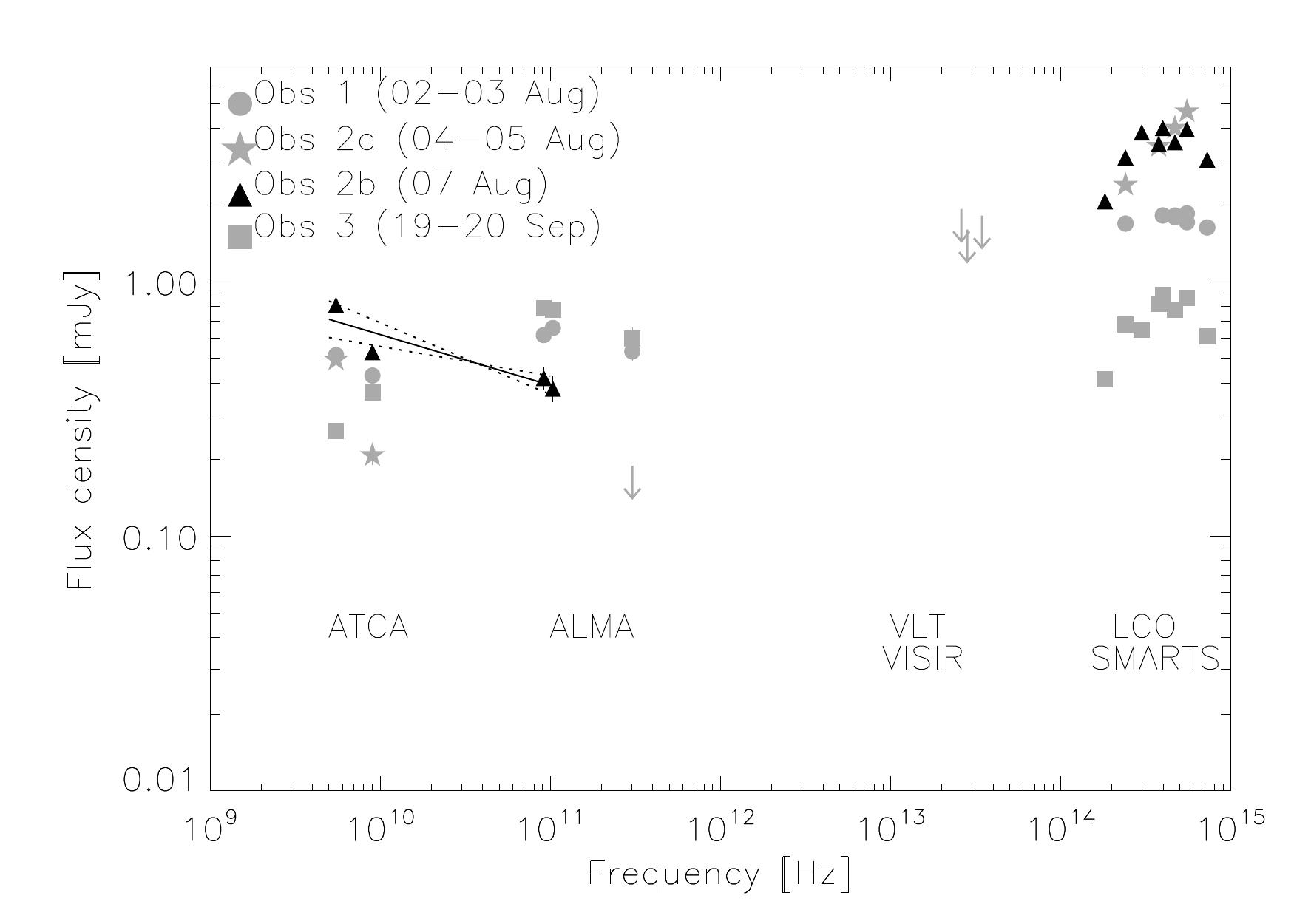}
\hspace{0.8cm}
\includegraphics[angle=0.0,width=0.35\textheight]{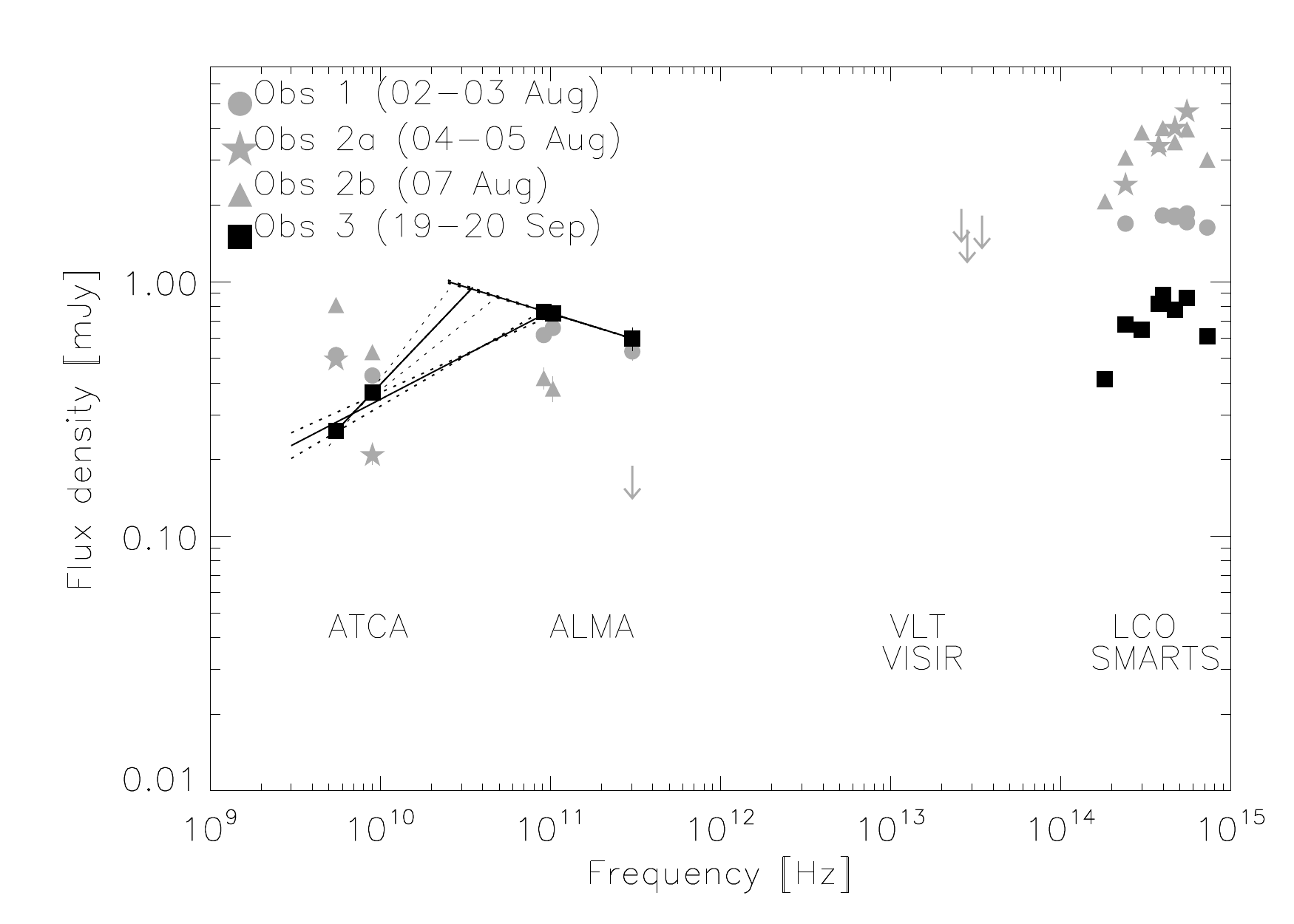}
\caption{Broad-band spectra for \srcone\ for obs~1--3. The arrows represent 3\,$\sigma$ upper limits. Obs~1, 2a, 2b and 3 are shown in black in the upper-left, upper-right, lower-left and lower-right panels, respectively. For obs~3 we show two possible fits for the first slope: considering only the ATCA points and considering the ATCA points together with the ALMA points. The jet frequency break could be anywhere between $\sim$30~GHz and 100~GHz depending on the solution taken. The jet break frequency evolves from $\sim$100~GHz in obs~1, to $\approxlt$5~GHz in obs~2a and 2b and back to 30--100~GHz in obs~3.}
\label{fig:sed}
\end{figure*}

We constructed broad-band spectra for the four pointed observations (see Fig.~\ref{fig:sed}). The radio to mm spectra for obs~1 and 3 are relatively similar in shape, with the spectral slope being positive from radio to mm frequencies, as expected from optically thick synchrotron emission from a jet. A spectral break occurs at frequencies of $\sim$100~GHz and the spectral slope 
becomes negative above that frequency. The mid-IR upper limit obtained for obs~1 is consistent with this scenario. In contrast, the spectra for obs~2a and 2b show a negative spectral slope from radio to mm frequencies, i.e. the peak of emission occurs at the lowest observed frequency, 5.5 GHz. Interestingly, obs~2b shows stronger flux densities compared to obs~2a, possibly indicating a jet re-brightening. The spectral slope is steeper below 9~GHz than between 9 and 97.49~GHz for obs~2b. Obs~2a could in principle also show two different spectral slopes below and above 9~GHz but we cannot determine if this is really the case due to the flux upper limit at 302.99~GHz.  The OIR spectra show a significant increase in flux from obs~1 to obs~2a/2b and the smallest flux is present in obs~3. 

We next fitted the spectral slopes for all the radio to mm spectra. For obs~1 and 3, a break at $\sim$100~GHz is apparent (see Fig.~\ref{fig:sed}). Therefore, we used the first four points/last three points to determine the spectral slope before/after the break. For obs~1, we obtain indices of 0.11\,$\pm$\,0.06 and --0.15\,$\pm$\,0.06 before and after the break, respectively, where $F_{\nu} \propto \nu^{\alpha}$ and $F_{\nu}$ is the flux density at frequency $\nu$. These indices are consistent with the spectrum being the result of optically thick/optically thin synchrotron
emission from a jet, before and after the break, respectively. Similarly, for obs~3, we obtain indices of 0.34\,$\pm$\,0.05 and -0.21\,$\pm$\,0.01 before and after the break at $\sim$100~GHz, respectively. For this  observation, the break could lie instead at frequencies as low as $\sim$30~GHz, if we consider the optically-thick slope traced by the ATCA points only. In contrast, for obs~1, the ATCA points trace a negative slope, which could indicate some structure in the broad-band spectra, unaccounted-for systematic uncertainties or time variability.  Also, the break could lie at frequencies higher than 100~GHz if we assume a steeper slope after the break. Specifically, to set a tentative upper limit on the frequency break for obs~1, we calculated the slope of the optically thin spectrum assuming an extension of the optically thick spectrum to 200 and 250~GHz. We obtained values of --0.6\,$\pm$\,0.3 and --1.4\,$\pm$\,0.7, respectively, for that slope. Therefore, we set a conservative upper limit of 250~GHz for the break frequency but note that this would already yield a very extreme slope, similar to that of obs~2a (see below), and is therefore unlikely.
For obs~2a we obtain a spectral slope of --1.76\,$\pm$\,0.25 considering only the 5.5 and 9~GHz detections. Considering the ATCA detections and the ALMA 3$\sigma$ upper limit we obtain an upper limit on the spectral slope of --0.19. For obs~2b we obtain a spectral slope of --0.20\,$\pm$\,0.08 between 5 and 97.49~GHz. Interestingly, for this observation the slope traced by the ATCA points only is significantly steeper than when fitting a single power law to both the ATCA and the ALMA points. Similarly to obs~1, this could indicate again some structure in the broad-band spectra. However, the different slopes at radio and mm frequencies could be simply explained by variability between the observations that are not strictly simultaneous.

\begin{table}
\begin{center}
\caption[]{Accretion state and radio to mm spectral indices. For obs~1 and 3, we obtained two jet spectral breaks per observation by taking into account different scenarios (see text). The two values are considered as lower and upper limits of the break for each observation. The true value very likely lies between these two extremes.}
\begin{tabular}{l@{\extracolsep{1mm}}l@{\extracolsep{1mm}}c@{\extracolsep{1mm}}c@{\extracolsep{1mm}}c@{\extracolsep{1mm}}}
\hline \noalign {\smallskip}
Obs. & State & Slope 1 & Slope 2 & Jet break \\
\hline \noalign {\smallskip}
1 & Hard/Intermediate & 0.11\,$\pm$\,0.06 & --0.15\,$\pm$\,0.06 & $\sim$100~GHz \\
1 & Hard/Intermediate & 0.11\,$\pm$\,0.06 & --1.4\,$\pm$\,0.7 & $\sim$250~GHz \\
2a & Soft & & --1.76\,$\pm$\,0.25 & $\approxlt$5~GHz  \\
2b & Soft & & --0.19\,$\pm$\,0.09 & $\approxlt$5~GHz  \\
3 & Hard/Intermediate & 0.70\,$\pm$\,0.17 & --0.21\,$\pm$\,0.01 & $\sim$30~GHz \\
3 & Hard/Intermediate & 0.34\,$\pm$\,0.05 & --0.21\,$\pm$\,0.01 & $\sim$100~GHz \\
\noalign {\smallskip} \hline 
\label{tab:indices}
\end{tabular}
\end{center} 
\end{table}

\section{Discussion}
\label{sec:discussion}
\subsection{Jet emission during soft states}

The evolution of the radio emission across outbursts was first systematically studied for \srcone\ by \citet{aqlx1:tudose09mnras} for the outbursts from 2002, 2004 and 2005. They reported a significant drop in radio fluxes above a bolometric flux of 5\,$\times$\,10$^{-9}$ erg cm$^{-2}$ s$^{-1}$ (corresponding to 10\% of the Eddington luminosity at a distance of 5.2~kpc) and suggested that such a drop could be analogous to the suppression of radio emission in BH XRBs in soft, bright, X-ray states. \citet{aql:miller-jones10apj} also observed a quenching of radio emission (3\,$\sigma$ upper limits of 0.17 and 0.08~mJy beam$^{-1}$ at 4.8 and 8.4 GHz, respectively) at bolometric fluxes larger than 10\% of Eddington during the November 2009 outburst (classified as S-type, \citet[][]{aqlx1:asai12pasj} or short-low, \citet{aqlx1:gungor14mnras}, respectively). 

In contrast, we detect radio emission during obs~2a and both radio and mm emission during obs~2b, at bolometric fluxes of 60 and 65\% Eddington, respectively, and with X-ray spectra consistent with a soft state. It is also remarkable that although the radio flux has dropped from obs~1 to 2a, during obs~2b we find a flux density of 816\,$\pm$\,35~$\mu$Jy at 5.5~GHz, the highest ever observed at this frequency. Moreover,  the spectra of obs~2a and 2b show a negative slope. Based on the monitoring of the 2009 outburst, \citet{aql:miller-jones10apj} showed that the spectral slope at radio frequencies was consistent with being flat outside of the hard state, i.e. consistent with optically thick emission, and suggested that the internal shock mechanism believed to produce optically thin transient radio ejecta was not active in \srcone. While a negative slope of --0.80\,$\pm$\,0.35 was detected between 5.25 and 7.45~GHz with the VLA during the 2013 outburst \citep[of long-high type,][]{aqlx1:gungor14mnras} at the hard-to-soft transition, the large flux uncertainty did not allow to confirm the presence of optically thin flares \citep{aql:miller-jones13atel5148}. Our observations 2a and 2b demonstrate the existence of radio to mm emission with a negative slope after the source has transited into the soft state and at luminosity levels of 60--65\% Eddington, well above the previous limits. 

Radio emission with negative slopes from BH transients during soft states has been detected in a handful of sources  \citep{corbel04apj,1630:neilsen14apjl} and interpreted in a scenario where the core jet is strongly suppressed and the emission is associated with optically thin emission from re-brightenings at shocks as the jet propagates away from the binary \citep{fender09mnras}. In such a scenario the emission should monotonically fade as the ejecta expand in the surrounding medium, unless new ejecta are launched. While the existence of a core jet during the soft state cannot be ruled out, it would then be difficult to explain why the observed emission is optically thin, as opposed to that of the hard state. Indeed, the spectral shape of the jet does not significantly change (only the intensity) in models invoking the existence of a core jet during soft states \citep{malzac14mnras,gx339:drappeau17mnras}. 
Our observations show that, similarly to BHs, NSs can also show radio to mm emission shortly after the transition to the soft state with a negative slope. Interestingly, obs~2a shows a  very extreme spectral index, albeit based on only two ATCA points. Such a steep slope, with an index $<$\,--1.5, has been only seldom previously observed \citep{fender09mnras,1659:horst13mnras} and at least in one case it has been associated to a spectral break at radio frequencies. While such indices seem too steep to be attributed to standard optically thin synchrotron, it might be that the emission evolves on different timescales at different frequencies or that the electron distribution has a sharp cutoff. 

Obs~2b shows a less negative spectral index than obs~2a. However, there seems to be some structure in the spectrum, namely the slope derived from the ATCA points only is much steeper (with an index of --0.87) than when the ATCA and ALMA points are considered. This could be due to variability between the ALMA and ATCA observations, taken 8 hours apart. 

Regardless of the spectral slope, it is clear that it is negative, similar to BHs and in contrast to the optically thick emission detected from persistent NSs in a steady soft state   \citep{1820:migliari04mnras,1820:diaz17aa}. This implies that two different types of jet emission may be present during soft states of NSs: optically thin emission in transient 
sources and optically thick emission in steady states of persistent sources. We note that in some BH sources, several radio flares from discrete ejecta have been seen to be launched within days of the hard to soft transition \citep[e.g. XTE J1859+226; Swift J1745--26;][]{brocksopp02mnras,fender09mnras,curran14mnras}. It is possible that because of the rapid transition of Aql X--1, and the sparse radio coverage, some radio flares may have been missed. Another possibility is that the flares are associated only to the brightest outbursts. While the 2011, 2013 and 2016 outbursts all belong to the ``long-high'' class based on the duration and maximum flux of the X-ray emission \citep{aqlx1:gungor14mnras}, the 2002 and the 2009 outbursts belong to a different, less bright, category. 

\subsection{Evolution of the jet spectral break along the outburst}

During the last years major effort has been devoted to determining what drives the jet ejection and reignition during spectral state transitions, whether jet ejections are always launched, and whether a weak compact jet might be present at all during soft states. As discussed above, the emission observed during soft states has been generally attributed to optically thin emission from jet ejecta. However, there is at least one BH, MAXI J1659--152, for which radio emission with a very steep negative slope has been detected in a soft state \citep{1659:horst13mnras} despite the absence of detections of ejecta with e-VLBI \citep{1659:paragi13mnras}. Moreover, similar optically thin emission has been also detected for the BH GX~339--4 at the reverse transition from soft to hard states before the appearance of an optically thick jet \citep{gx339:corbel13bmnras}. In both cases, the break frequency is below $\sim$10~GHz and seems to indicate that the negative slope spectra may be also a key to the mechanism of how jets switch on and off. Perhaps the clearest indication of this is the continuous evolution of the spectral break towards lower frequencies and back to higher frequencies in the BH \maxi\ as the outburst evolves from a hard to an intermediate state and back \citep{1836:russell14mnras}. At least in this case, the evolution is continuous and even if the transition to a soft state never occurs optically thin emission is observed down to a break frequency of $\approxlt$30~GHz. Moreover, the break frequency is directly correlated to the hardness of the X-ray spectrum. Interestingly, such a correlation is also found when the comparison is extended to super-massive BHs \citep{koljonen15apj}. 

The few break frequencies detected until now in NSs in hard states lie in the IR band \citep{0614:migliari10apj,baglio16aa,1820:diaz17aa}. Our observations show for the first time the temporal evolution of the break frequency for a NS. The frequency of the break in obs~1 is at a lower frequency than that of previously detected breaks during hard states. While we cannot discard that the break lies at this low frequency in hard states well before the transition, it seems plausible that the low frequency is a result of the evolution at the start of the spectral transition. This observation highlights the importance of truly broad-band coverage for jet studies since based on the radio spectra alone we would have concluded that the spectral break was already below 5~GHz at this stage. Since the ALMA observations were performed after the ATCA ones, we can discard that the break moved towards lower frequencies between the ATCA and the ALMA observations. 

Following the transition to a soft state, both obs~2a and 2b show a steep spectrum with no apparent break in the frequency range observed. Therefore, we conclude that if the compact jet still exists, the break must be below $\sim$5~GHz. The re-brightening during obs~2b could be attributed to an enhanced episode of mass channelled into the jet. For this to happen, either a core jet should still be present during obs~2a or the core jet should switch off and on between obs~2a and 2b. Alternatively, the re-brightening could be due to a new discrete ejection, a shock between ejected material and the ISM, or even with the interaction of two ejecta at different speeds. 

During obs~3, the jet break is detected at $\sim$\,30--50~GHz (if we consider only the ATCA points) or $\sim$100~GHz (if we fit the ATCA and ALMA low frequency points together, see Fig.~\ref{fig:sed}). Should the break frequency be lower at the same hardness at the outburst decay, this could be due to the reduced mass accretion rate \citep{heinz03mnras,falcke04aa}. However, we cannot determine if this is the case from these observations. Interestingly, the flux densities detected at mm frequencies are the highest from all observations indicating that the mm emission is not directly correlated with the soft or hard X-ray flux and that a powerful jet can be present during the decay of the outburst.  
After obs~3, we would expect that the jet evolves by reigniting at higher frequencies up to OIR \citep[e.g.][]{gx339:corbel13bmnras}. However, for \srcone, the contribution of the jet at OIR frequencies is probably not dominant (if existent at all) during hard states, as opposed e.g. to the persistent neutron star 4U~1728--34 \citep{1820:diaz17aa} since we do not see any significant re-brightening at such frequencies during the decay. Alternatively the jet could have decayed too quickly to be detectable at any frequency after obs~3. 

\begin{figure*}[ht]
\includegraphics[angle=0.0,width=0.48\textwidth]{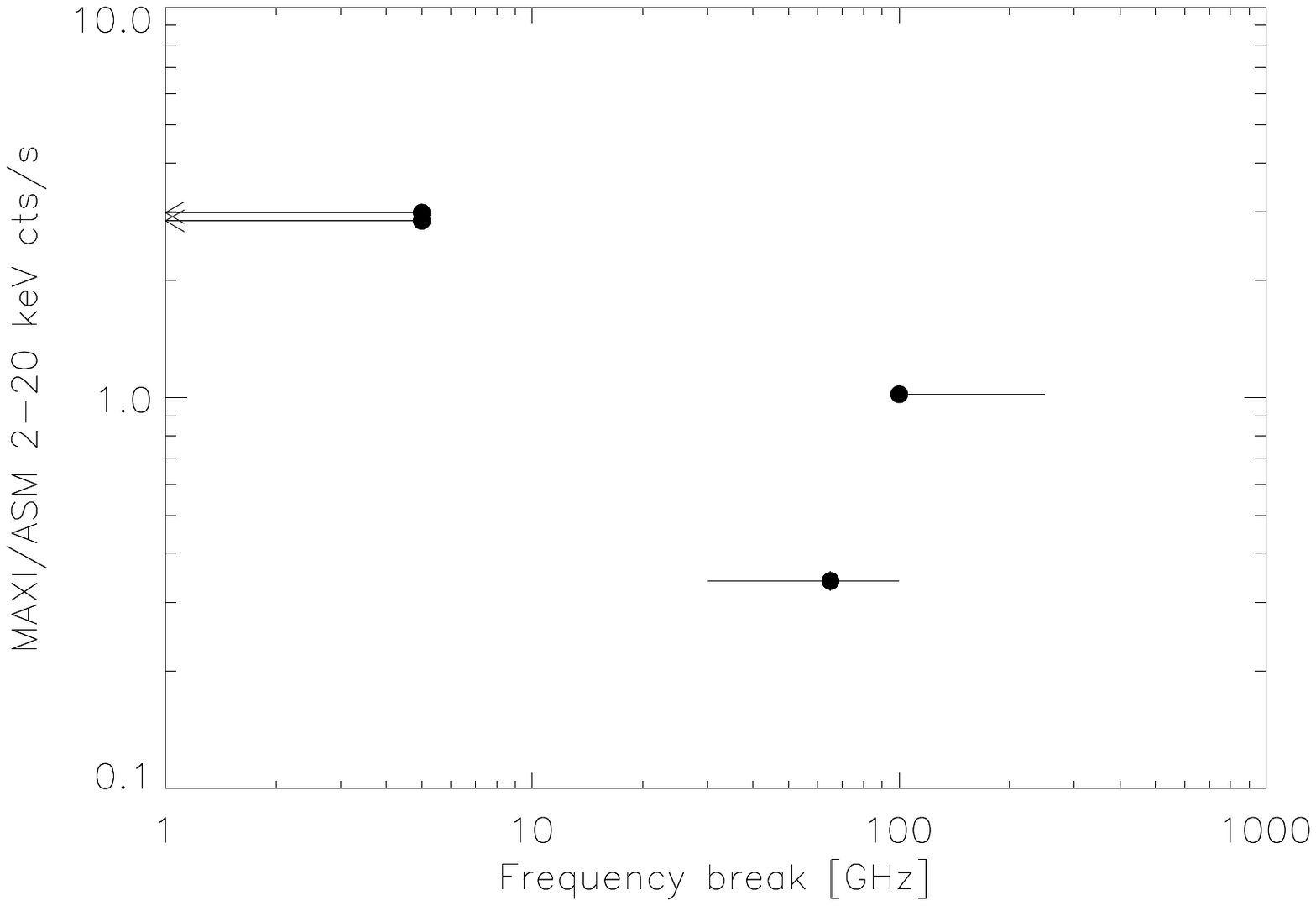}
\includegraphics[angle=0.0,width=0.48\textwidth]{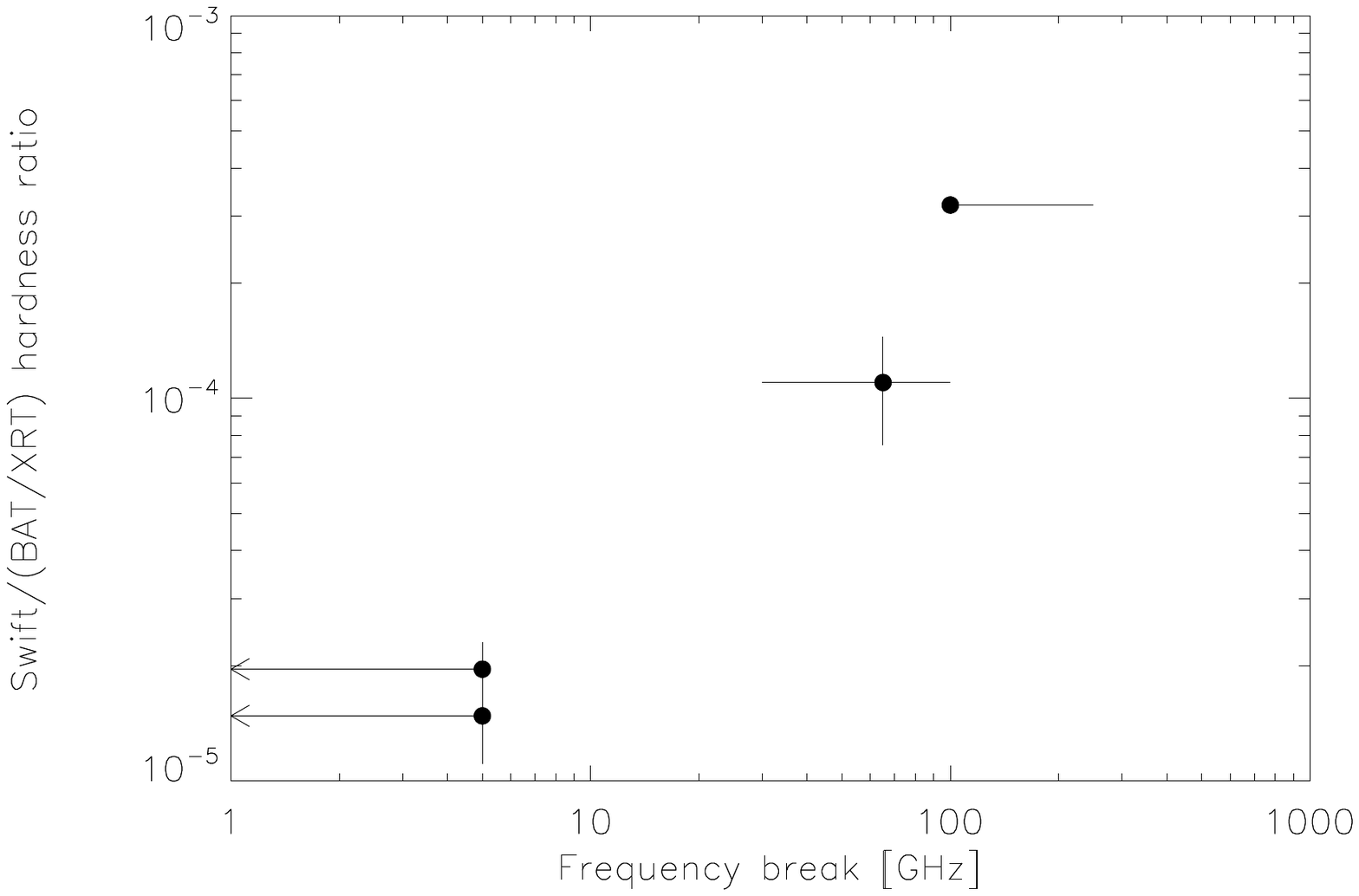}
\vspace{-5cm}
\caption{Frequency of the jet spectral break versus the MAXI/ASM X-ray count rate (left) and the $\swift$/BAT/XRT HR (right) for obs 1--3.}
\label{fig:break_hr}
\end{figure*}

Finally, we explored if the break frequency could be directly correlated to the hardness of the X-ray spectrum, as it has been recently found for BHs \citep{1836:russell14mnras, koljonen15apj}. Fig.~\ref{fig:break_hr} shows the break frequencies obtained for the four multi-wavelength observations versus the MAXI/ASM 2--20 keV count rate and the $\swift$/BAT/XRT HR. For obs~1/2a, we use the flux/HR during days 13/15, respectively, since these are the days when the $\swift$/XRT and ALMA observations were performed and note that at least the ATCA observation on day 12 may have happened during a slightly harder state than the corresponding $\swift$/XRT and ALMA  ones. For obs~3, we use the flux/HR during day 60 since all observations were performed on that day except the ALMA band 7 one, which was performed very close, on day 61.04. Similarly to what was observed for BHs, the frequency of the break increases with hardness of the X-ray spectra. While a Spearman rank test yields a positive correlation of 90$\%$ between the jet break frequency and the hardness of the X-ray spectrum, such a correlation still has a chance probability $>$\,5$\%$ given the low number of degrees of freedom (2). Thus, more data are needed to confirm this correlation in \srcone. 
A correlation of the break frequency with the X-ray flux is not observed. A caveat regarding this comparison is that in NSs the X-ray spectrum shows a significant contribution of the NS surface or boundary layer. Therefore, an interpretation of 
the dependence of the frequency break as a function of spectral components such as the accretion disc or hot corona is more difficult than for BHs. 

\section{Conclusions}

We observed the NS \srcone\ across the hard-to-soft transition and during the decay of its 2016 outburst from radio to X-rays. 
We detected radio emission in all cases, which comprise both hard/intermediate and soft accretion states. Millimetre emission was also detected in three out of four observations, including one soft state. 
The broad-band spectra from radio to mm frequencies evolve from having a spectral break from optically thick to thin emission at $\sim$100~GHz during hard/intermediate states to below $\sim$5~GHz during soft states assuming the compact jet persists. We do not find any correlation of the X-ray flux with the spectral break. In contrast a correlation of the break frequency with X-ray spectral hardness is possible, like that discovered in BH transients. These observations show that the processes at play in BH jets are also present in NS transients and highlight the importance of simultaneous multi-wavelength coverage of XRB outbursts. 

\begin{acknowledgements}  
This paper makes use of the following ALMA data:
   ADS/JAO.ALMA\#2015.1.00734.T. ALMA is a partnership of ESO (representing
   its member states), NSF (USA) and NINS (Japan), together with NRC
   (Canada) and NSC and ASIAA (Taiwan), in cooperation with the Republic of
   Chile. The Joint ALMA Observatory is operated by ESO, AUI/NRAO and NAOJ.
   The Australia Telescope Compact Array is part of the Australia Telescope National 
   Facility which is funded by the Australian Government for operation as a National Facility managed by CSIRO. D. A. 
   acknowledges support from the Royal Society. J. C. A. M.-J. is the recipient of an Australian Research Council Future Fellowship (FT140101082).
   D. M. R. thanks Mario van den Ancker at ESO for help with the preparation and flux calibration of the VLT/VISIR observations. 
   The Faulkes Telescope Project is an education partner of Las Cumbres Observatory. The Faulkes Telescopes are maintained and operated by LCO.
\end{acknowledgements}


\bibliographystyle{aa}
\bibliography{alma_aqlx-1}

\end{document}